\documentclass[tightenlines,eqsecnum,floats,aps,amsmath,amssymb,nofootinbib,prd,showpacs,notitlepage]{revtex4-1}
\pdfoutput=1 
\usepackage{amsmath,amssymb,amsfonts}
\usepackage{graphicx}
\usepackage{enumerate} 
\usepackage{colordvi} 
\usepackage{bm}
\usepackage{epsfig}
\usepackage{float}
\usepackage{verbatim}
\usepackage{subfig}
\usepackage{xcolor}

\usepackage[colorinlistoftodos]{todonotes}
\usepackage[colorlinks=true, allcolors=blue]{hyperref}

\newcommand{\bfk}{\mbox{\boldmath$k$}}

\newcommand{\mnras}{MNRAS}

\newcommand{\aap}{AAP}

\begin{document}
\title{Assessing non-linear models for galaxy clustering III: theoretical accuracy for Stage IV surveys}


\vfill
\author{Benjamin Bose}
\affiliation{Departement de Physique Theorique, Universite de Geneve, 24 quai Ernest Ansermet, 1211 Geneve 4, Switzerland}
\author{Kazuya Koyama}
\affiliation{Institute of Cosmology \& Gravitation, University of Portsmouth, Portsmouth, Hampshire, PO1 3FX, UK}
\author{Hans A. Winther}
\affiliation{Institute of Cosmology \& Gravitation, University of Portsmouth, Portsmouth, Hampshire, PO1 3FX, UK}
\affiliation{Institute of Theoretical Astrophysics, University of Oslo, 0315 Oslo, Norway}
\bigskip

\bigskip
\vfill
\date{\today}
\begin{abstract}
We provide in depth MCMC comparisons of two different models for the halo redshift space power spectrum, namely a variant of the commonly applied Taruya-Nishimichi-Saito (TNS) model and an effective field theory of large scale structure (EFTofLSS) inspired model. Using many simulation realisations and Stage IV survey-like specifications for the covariance matrix, we check each model's range of validity by testing for bias in the recovery of the fiducial growth rate of structure formation. The robustness of the determined range of validity is then tested by performing additional MCMC analyses using higher order multipoles, a larger survey volume and a more highly biased tracer catalogue. We find that under all tests, the TNS model's range of validity remains robust and is found to be much higher than previous estimates. The EFTofLSS model fails to capture the  spectra for highly biased tracers as well as becoming biased at lower wavenumbers when considering a very large survey volume. Further, we find that the marginalised constraints on $f$ for all analyses are stronger when using the TNS model.
\end{abstract}

\maketitle


\section{Introduction}\label{intro}
Future spectroscopic galaxy surveys will have the power to probe the growth rate of cosmological structure, $f$, to an unprecedented level of precision. The measurement of the anisotropy in the galaxy distribution, the so called redshift space distortions (RSD) \cite{Hamilton:1997zq}, is one such way to get a measure of this, and  will tell us a lot about gravity and cosmology. This comes with the caveat that we apply an accurate theoretical model to the data.  If the model we apply is not sufficiently accurate then the value of $f$ we infer from the data will not be the 'true' value, and our picture of nature will be biased. 
\newline
\newline
This caveat becomes very important in the context of stage IV surveys such as Euclid\footnote{\url{www.euclid-ec.org}} \cite{Laureijs:2011gra} and the Dark Energy Spectroscopic Instrument (DESI)\footnote{\url{www.desi.lbl.gov}} \cite{Aghamousa:2016zmz}.  The issue here is that the data will come with very small observational errors which will greatly penalize small inaccuracies in our modelling. This then requires that applied models be heavily scrutinized and tested before being applied to real observational data and drawing conclusions about the universe. 
\newline
\newline
The two point correlation function, or power spectrum, has been the observable commonly used in past data sets \cite{Blake:2011rj,Reid:2012sw,Macaulay:2013swa,Beutler:2013yhm,Gil-Marin:2015sqa,Simpson:2015yfa,Beutler:2016arn}. As it uses only galaxy pairs, it can be measured with relatively high statistical significance, a significance that is enhanced as we move to smaller galaxy separations. For this reason, modeling the smaller scales becomes an attractive means of tightening constraints on growth. But as mentioned, this needs to be done carefully and any potential power spectrum models must be well validated. These models usually come with additional degrees of freedom (dof) that are necessary to model largely unknown physics, such as non-linear matter dynamics or the bias between galaxy and dark matter distributions. These so called nuisance parameters degrade  constraints on cosmology and gravity in data analyses as they are marginalised over. Thus, a model whose nuisance parameters are few and non-degenerate with cosmological ones is strongly preferred.
\newline
\newline
In a previous work \cite{Bose:2019psj}, we identified two prominent and competing semi-perturbative models for the redshift space galaxy power spectrum. Namely, the TNS model \cite{Taruya:2010mx} with a Lorentzian damping prefactor and an effective field theory of LSS (EFTofLSS) \cite{Baumann:2010tm,Carrasco:2012cv} inspired model. Both these models attempt to model the quasi non-linear regime of structure formation in different ways and both were shown to do well in reproducing the measured halo power spectrum from COLA \cite{Tassev:2013pn,Izard:2015dja, Howlett:2015hfa, Valogiannis:2016ane, Winther:2017jof} simulations when $f$ was fixed to its fiducial value. In that work we were concerned with determining the degeneracies of each model's dofs as well as forecasting the model's constraint on $f$ in the context of a future spectroscopic survey. We also obtained realistic constraints in a limited Markov Chain Monte Carlo (MCMC) analysis.
\newline
\newline
In this work we go a step further in testing how robust each of these model's accuracy is by making use of various data sets in the setting of a real data analysis. In particular, we use a large suite of high resolution COLA simulations to test the accuracy of the TNS and EFTofLSS model. We do this by letting the growth rate of structure and all additional model dof to vary in further MCMC analyses. We then identify where each model fails to reproduce the fiducial growth rate and compare their respective constraints. 
\newline
\newline  
This paper is organized as follows: In Sec.~\ref{theory} we present the biased tracer RSD models. In Sec.~\ref{sims} we present the simulations we use and our primary MCMC analysis. In Sec.~\ref{results} we present the results from additional analyses testing the robustness of the models to different data sets and errors. In Sec.~\ref{summary} we summarize our findings and conclude. 


\section{Theoretical Models}\label{theory}
We will begin by presenting the two biased tracer RSD models. These were described and studied in \cite{Bose:2019psj} and we refer the reader to this paper for more explicit information on the exact formulas. 
\newline
\newline
The first is the TNS RSD model \cite{Taruya:2010mx} combined with the tracer bias model of McDonald and Roy \cite{McDonald:2009dh}. The model is given by 
 \begin{align}
 P^S_{TNS}(k,\mu) =& D_{\rm FoG}(\mu^2 k^2 \sigma_v^2)\Big[ P_{g,\delta \delta} (k,b_1,b_2,N) + 2 \mu^2 P_{g,\delta \theta}(k,b_1,b_2) +  \mu^4 P_{\theta \theta}^{\rm 1-loop} (k) \nonumber \\ & \qquad \qquad \qquad \qquad + b_1^3A(k,\mu) + b_1^4B(k,\mu) +b_1^2 C(k,\mu)  \Big], 
 \label{redshiftps}
 \end{align} 
where the superscript S denotes the power spectrum in redshift space, $\mu$ is the cosine of the angle between $\bfk$ and the line of sight and $P_g$ are the 1-loop galaxy power spectra with the bias model of \cite{McDonald:2009dh} implicitly included. The logarithmic growth rate of structure $f$ is also implicit (again see \cite{Bose:2019psj}).  The real space power spectra are all constructed within standard Eulerian perturbation theory  at the 1-loop level \footnote{ Note that within the loop integrals we parametrise the integrated wave number as $k'=kr$ and then take a UV cut-off of $r=10$. We have found that the integrals are insensitive to this choice at or above the chosen value. See \cite{Makino:1991rp} for a discussion on this issue.} , $A$,$B$ and $C$ are perturbative RSD correction terms \cite{Taruya:2010mx}, while the prefactor, $D_{\rm FoG}$, is phenomenological and here takes a Lorentzian form 
\begin{equation}
    D_{\rm FoG}^{\rm Lor}(k^2\mu^2 \sigma_v^2) = \frac{1}{1 + (k^2\mu^2 \sigma_v^2)/2},
\end{equation}
where $\sigma_v$ is a free parameter and represents the velocity dispersion of the tracers. We again refer the reader to \cite{Bose:2019psj} for the formulas for the perturbative components of the model, along with the explicit dependency on the independent free bias parameters $\{b_1,b_2,N\}$, where $b_1$ is the linear bias, $b_2$ is the second order bias term and $N$ is a stochasticity term.
\newline
\newline 
We remark that this model is very similar to the model chosen for the BOSS analysis \cite{Beutler:2016arn} and has been very well studied and successful in reproducing simulation measurements. It has also shown robustness when considering alternative theories of gravity (see for example \cite{Bose:2016qun,Bose:2017jjx}). The full set of {\bf nuisance parameters} in this model is $\{ \sigma_v, b_1, b_2, N \}$.
\newline
\newline
The second model we consider is one based on the EFTofLSS prescription for the redshift space dark matter spectrum (see \cite{delaBella:2017qjy} for example). To this we introduce the tracer bias model of \cite{McDonald:2009dh} (same as used in Eq.~\ref{redshiftps}) as well as use a resummation scheme \cite{Vlah:2015zda,delaBella:2017qjy} \footnote{ See \cite{Matsubara:2007wj,Senatore:2014vja,Lewandowski:2015ziq} for alternative schemes that have been shown to be equivalent to the one adopted here in \cite{Bose:2018orj}.} for the dark matter 1-loop power spectra only as these are the leading contributions to the  oscillations associated with baryon acoustic oscillations when fitting the counter terms. Note that we do not do a full resummation of the redshift space spectrum as in \cite{delaBella:2017qjy}, but the impact of resummation has also been shown to have a low impact on the best-fit analysis conducted in \cite{delaBella:2018fdb} where they consider halos in redshift space. The expression is given as 
\begin{align}
P^{S}_{eft}(k,\mu) = & \{1-(D_1^2f^2k^2\mu^2 \tilde{\sigma}^2_{v})\} \Big[ P_{g,\delta \delta} (k) + 2  \mu^2 P_{g,\delta \theta}(k) +  \mu^4 P^{1-{\rm loop}}_{\theta \theta} (k) \Big]  \nonumber \\ 
& \qquad \qquad \qquad \qquad  + b_1^3A(k,\mu) + b_1^4B(k,\mu) + b_1^2C(k,\mu) \nonumber \\ 
&-2D_1^2P_L(k) k^2 \Big[c^2_{s,0} + {c}^2_{s,2} \mu^2 + {c}^2_{s,4} \mu^4 + \mu^6 (f^3 {c}^2_{s,0} - f^2 {c}^2_{s,2} + f {c}^2_{s,4})\Big],
\label{redshiftps2}
\end{align}
where $c_{s,i}$ are the sound speed parameters of EFTofLSS, $D_1$ is the linear growth factor, $P_L(k)$ is the primordial power spectrum \footnote{Produced using CAMB \cite{Lewis:2002ah} for example.} and 
\begin{equation}
\tilde{\sigma}^2_{v} = \frac{1}{6\pi^2} \int dq P_L(q).
\end{equation}
The {\bf nuisance parameters} of this model are $\{ b_1,b_2,N, {c}_{s,0},{c}_{s,2},{c}_{s,4} \}$ which is an additional 2 over the TNS approach described by Eq.~\ref{redshiftps}. A slight variant of Eq.~\ref{redshiftps2} was found to be well motivated in \cite{delaBella:2018fdb} through a Bayesian criterion which accounts for number of free parameters of the model as well as fits to simulations. On the other hand, the proper treatment of bias in EFTofLSS has been presented in \cite{Perko:2016puo} but comes with 10 nuisance parameters (compared to the 6 of Eq.~\ref{redshiftps2}). Given this, it is very unclear whether this model would be favoured over the ones presented here. 
\newline
\newline 
In Eq.~\ref{redshiftps} and Eq.~\ref{redshiftps2} cosmological parameter dependence enters through the primordial power spectrum $P_L(k)$ with the parameter $\sigma_8$ \footnote{$\sigma_8^2 = \frac{1}{2\pi^2}\int |W(kR)|^2P(k)k^2{\rm d}k$ governs the amplitude of density perturbations at $R = 8$Mpc/$h$ where $W(x) = \frac{3(\sin(x)-x\cos(x))}{x^3}$ is the tophat window function.} being completely degenerate with the linear growth factor $D_1$. In our analysis we fix $D_1$ or equivalently $\sigma_8$. Further we fix a fiducial cosmology and in doing so any results made on the robustness or range of validity are consequently conservative. Any results on constraining power are conversely optimistic as we do not marginalize over cosmology.


\section{Simulations}\label{sims}
The simulations used in this paper were created using a modified version \cite{Winther:2017jof} of the \texttt{L-PICOLA} code \cite{Howlett:2015hfa}. These simulations use the fast approximate COmoving Lagrangian Acceleration (COLA) method \cite{Tassev:2013pn}. We created a set of $35$ realisations of a $\Lambda$CDM cosmology defined by $\Omega_m = 0.307, \Omega_b = 0.0482, h = 0.678, \sigma_8 = 0.823, n_s = 0.961$. The simulations had $N = 1024^3$ particles in a box of $B = 1024\,\text{Mpc}/h$ with $N_{\rm grid} = 3000^3$ grid-cells. We also ran a full {\it N}-body simulation using the \texttt{RAMSES} \cite{2002A&A...385..337T} code that had the same initial condition as one of the COLA simulations which allowed us to check the accuracy of our results. These results can be found in Appendix~\ref{app:colanbodycomp}.
\newline
\newline
From the simulations we computed halo catalogs using a friends-of-friends (FOF) algorithm\footnote{The halo finder \texttt{MatchMaker} has been included in the \texttt{MGPICOLA} code used in this paper. See https://github.com/damonge/MatchMaker and  https://github.com/HAWinther/MG-PICOLA-PUBLIC}. The halo catalogs were then trimmed based on a number density criterion (we considered samples with $n = 10^{-3}$ and $10^{-4} (\text{Mpc}/h)^{-3}$) giving us the mock data from which we computed the RSD multipoles and the real-space power spectrum.
\newline
\newline
 PICOLA multipoles are measured using the distant-observer approximation. That is, we assume the observer is located at a distance much greater then the box size ($r\gg 1024 \, \mbox{Mpc}/h$), so we treat all the lines of sight as parallel to the chosen Cartesian axes of the simulation box. Next, we use an appropriate velocity component ($v_x, v_y$ or $v_z$) to displace the position of a matter particle or dark matter halo to put it into redshift space. We then average over three line-of-sight directions. We further average over the $35$ PICOLA realisations. These are calculated at both $z=0.5$ and $z=1$. 


\section{Results}\label{results}

Our aim here is to test the accuracy of each of the models outlined in Sec.~\ref{theory} and their ability to model the non-linear regime. To do this we perform a fit to the simulated data. First we perform an angular decomposition of $P(k,\mu)$ in terms of the multipole moments. This is what is commonly done in real data analyses \cite{Beutler:2013yhm,Beutler:2016arn}. These multipoles can be defined as 
\begin{equation}
P_\ell^{(S)}(k)=\frac{2\ell+1}{2}\int^1_{-1}d\mu P^{S}(k,\mu)\mathcal{P}_\ell(\mu),
\end{equation}
where $\mathcal{P}_\ell(\mu)$ denote the Legendre polynomials and $P^{S}(k,\mu)$ is given by Eq.~\ref{redshiftps} or Eq.~\ref{redshiftps2}. For the majority of our analyses we utilize the monopole ($\ell=0$) and quadrupole ($\ell=2$). The inclusion of the hexadecapole is known to significantly restrict the range of validity/applicability of the models considered. This partly comes from the fact that each multipole has a different damping scale and so we do not expect a model with less RSD parameters than the number of multipoles, to be able to capture all of them accurately. Higher order multipoles have finer features as a function of angle and so may also pick out inaccuracies in the full anisotropic spectrum. Since the information gain is roughly proportional to each independent $k$ one can access, a restriction in range results in a reduction in valuable cosmological information. So, since the monopole and quadrupole contain most of the RSD information we first perform fits using only $P_0$ and $P_2$ and then introduce $P_4$ but restrict the scales to which we compare it to the data. This is what was done in the BOSS analysis \cite{Beutler:2016arn}. An analysis using this procedure is outlined in a later subsection.  
\newline
\newline
Using the multipoles we perform a large number of MCMC analyses on the simulation data in order to test the model's recovery of the fiducial growth rate $f$ at various inclusion of scales. We vary all model nuisance parameters in these analyses as well as $f$, imposing the following flat priors $\sigma_v,c_{s,i}^2,b_1 \geq 0$. 
\newline
\newline
We model our log likelihood using the $\chi^2$ statistic 
\begin{equation}
-2 \ln(\mathcal{L}) = \sum_{k=k_{\rm min}}^{k_{\rm max}} \sum_{\ell,\ell'=0,2(,4)} \left[P^{S}_{\ell,{\rm data}}(k)-P^{S}_{\ell,{\rm model}}(k)\right] \mbox{Cov}^{-1}_{\ell,\ell'}(k)\left[P^{S}_{\ell',{\rm data}}(k)-P^{S}_{\ell',{\rm model}}(k)\right],
\label{covarianceeqn}
\end{equation}
where $\mbox{Cov}_{\ell,\ell'}$ is the covariance matrix between the different multipoles, $k_{\rm max}$ is the smallest physical scale used in the analysis and conversely $k_{\rm min} = 0.006 \, h/{\rm Mpc}$ is the largest physical scale. We apply linear theory to model the covariance matrix between the multipoles (see Appendix C of \cite{Taruya:2010mx} for details). This has been shown to reproduce {\it N}-body results up to $k\leq 0.3h/\mbox{Mpc}$ at $z=1$ \cite{Taruya:2010mx}. Again, for the majority of our analyses we only consider $\ell =0,2$ since these have the largest signal and contain most of the RSD information. We do consider the impact of $\ell=4$ in a later subsection.
\newline
\newline
We wish to work within the context of stage IV surveys. Therefore, we will concentrate our analysis on $z=1$ which will be a key redshift targetted by the Euclid mission \cite{Amendola:2016saw}. Further we use planned tracer number density, $n = 10^{-3} h^3/{\rm Mpc}^3$ and a realistic observational volume\footnote{Note that this depends on the bin-width, and here the volume chosen corresponds to a bin width of $\Delta z \sim 0.1$.} of $V=4 {\rm Gpc}^3/h^3$ \cite{Aghamousa:2016zmz,Majerotto:2012mf,Amendola:2016saw} in our covariance matrix in Eq.~\ref{covarianceeqn} \footnote{Note we also assume a linear bias in the covariance as determined by the simulation halo catalogs, i.e. the ratio of $P_{\delta \delta}$ to $P_{\delta h}$. For the $n=10^{-3} h^3/{\rm Mpc}^3$ catalog we take $b_1=2.03$ for $z=1$ and $b_1=1.49$ at $z=0.5$.}. This volume and number density are also reflected in our halo catalogues and unless otherwise stated, these will be the default parameters of the analysis. 
\newline
\newline
We will also provide supplementary analyses which will consider the hexadecapole, a lower redshift where non-linear structure formation is greater, a different catalog of halos with a lower number density and the impact of having a larger survey volume which will better reflect the combined power of the whole survey over many bins.


\subsection{Dark Matter} 
\begin{figure}[H]
\centering
  \includegraphics[width=8cm,height=8cm]{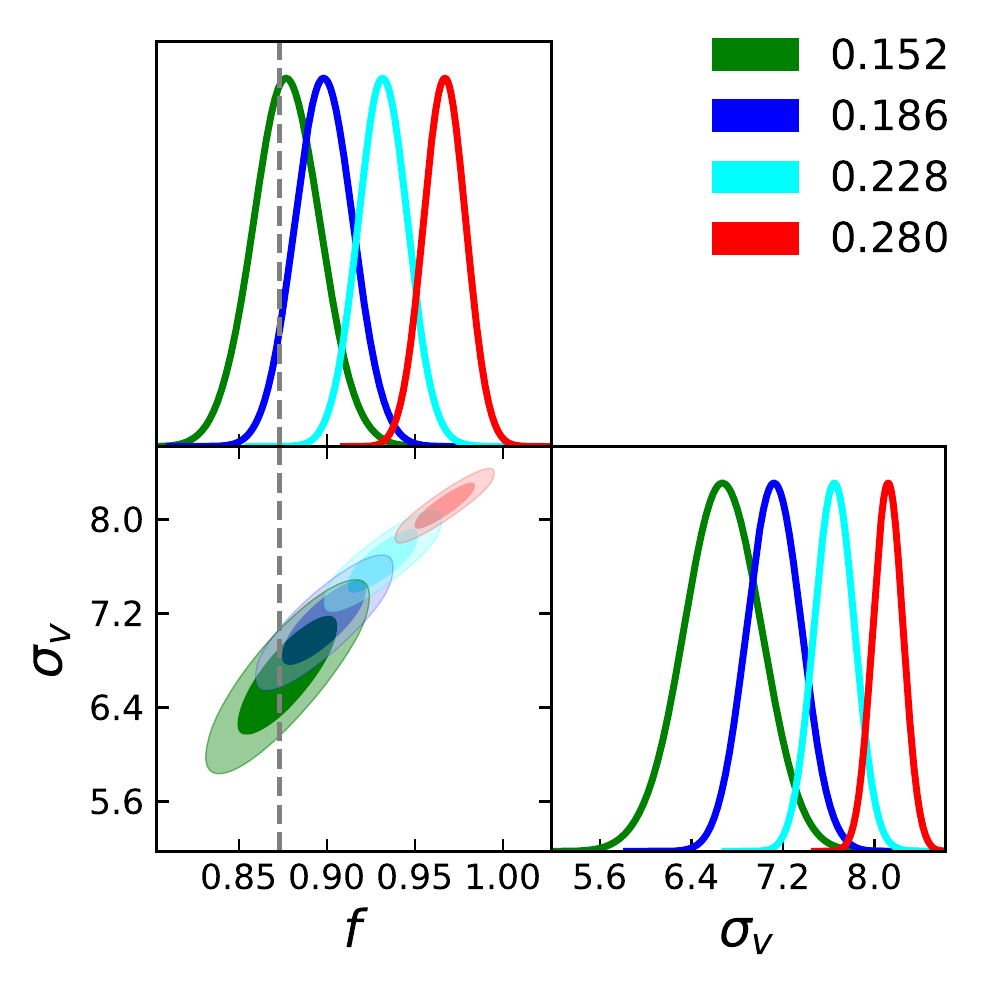}  \\  
  \includegraphics[width=15cm,height=12cm]{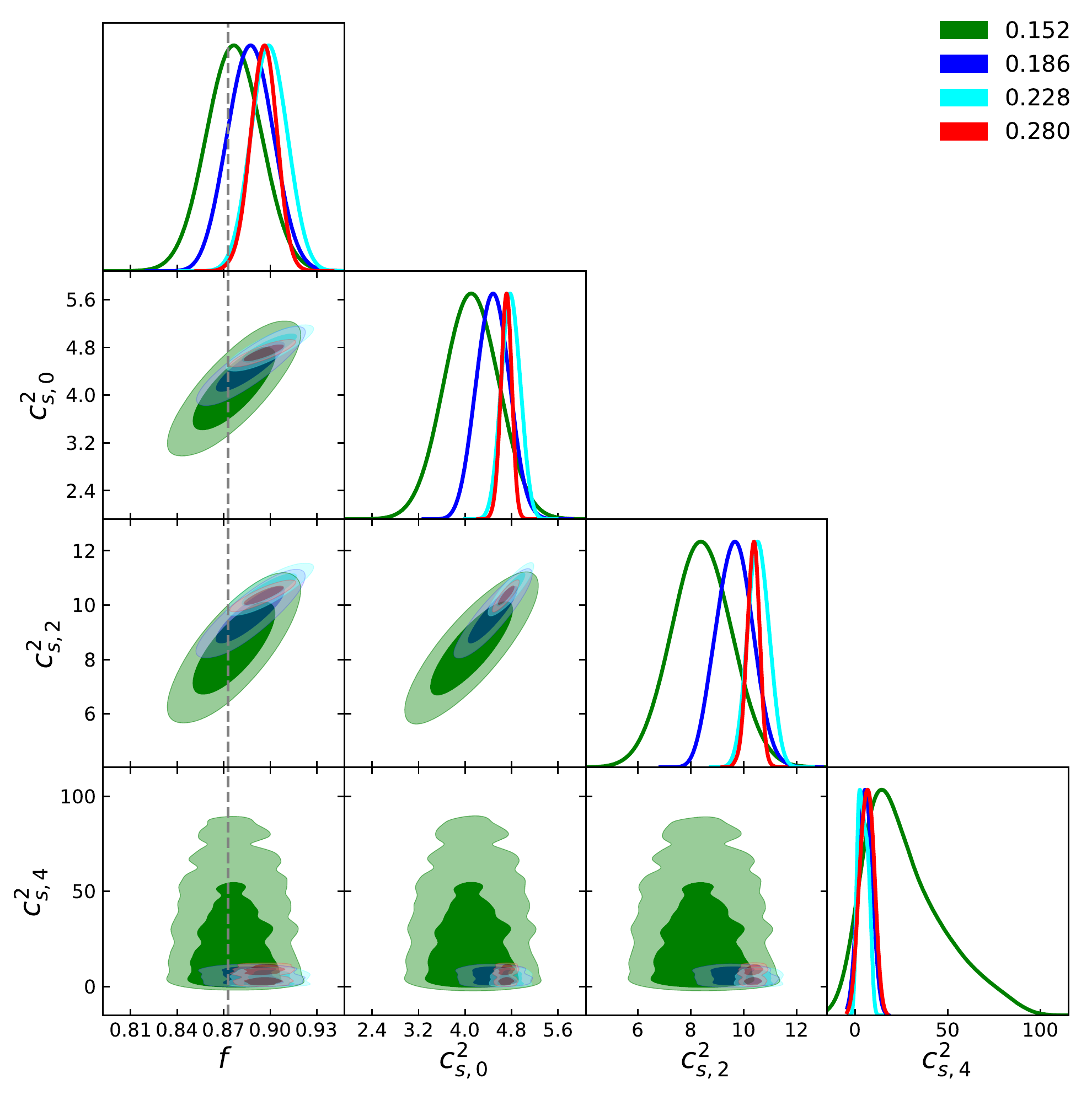} 
  \caption[CONVERGENCE]{ Redshift space dark matter results at $z=1$. { \bf Top:} The $1\sigma$ and $2\sigma$ confidence contours for the TNS model for varying $k_{\rm max} \in [0.152,0.280]h/{\rm Mpc}$. {\bf Bottom:} The $1\sigma$ and $2\sigma$ confidence contours for the EFTofLSS model for varying $k_{\rm max} \in [0.152,0.280]h/{\rm Mpc}$. The fiducial value of $f$ is denoted by a dashed line and only $P_0$ and $P_2$ were used in the analyses. }
\label{dm1}
\end{figure}

\begin{figure}[H]
\centering
  \includegraphics[width=12cm,height=8cm]{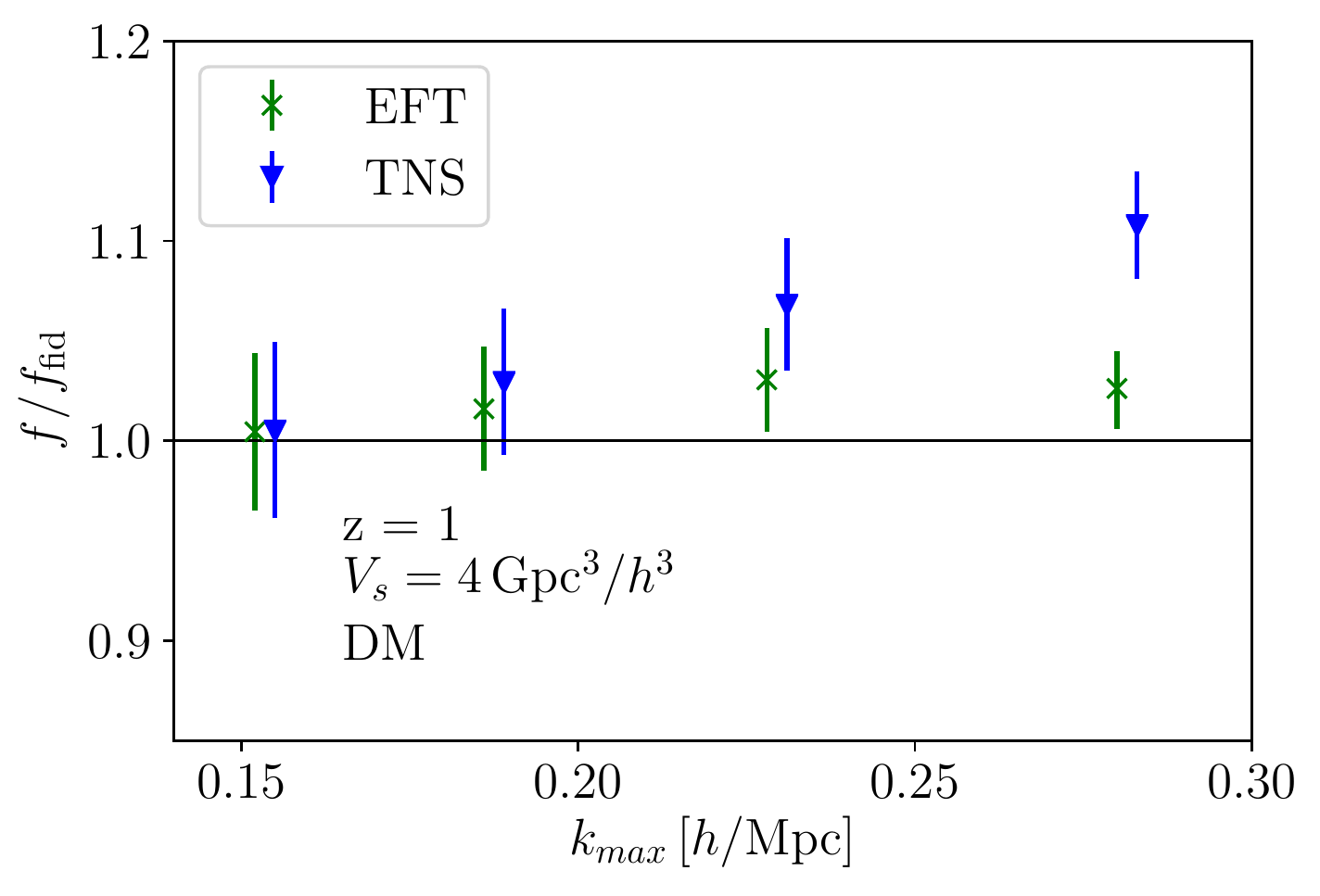}  
  \caption[CONVERGENCE]{Redshift space dark matter results at $z=1$. The mean value of $f/f_{\rm fiducial}$ as a function of $k_{\rm max}$ using the TNS (blue triangles) and EFTofLSS (green crosses) models with the marginalised $2\sigma$ error bars. Only $P_0$ and $P_2$ were used in the analyses.}
\label{dm2}
\end{figure}

Before considering halos, we look at the dark matter distribution  as it will be instructive to understand the capabilities of the pure RSD models, i.e. setting $b_1=1$ and $b_2 = N = 0$ in Eq.~\ref{redshiftps} and Eq.~\ref{redshiftps2}. Again, this is done at $z=1$ using $V=4 {\rm Gpc}^3/h^3$. Fig.~\ref{dm1} shows the two-dimensional posterior distributions for the TNS and EFTofLSS models when we only consider dark matter for various $k_{\rm max}$ used in Eq.~\ref{covarianceeqn} while Fig.~\ref{dm2} shows the marginalised constraints on $f$ as a function of $k_{\rm max}$. We clearly see both models become biased  with a $2\sigma$ criterion in their recovery of the fiducial value of $f$ at around $k_{\rm max} = 0.186h/{\rm Mpc}$. The TNS model diverges quickly as we include even smaller scales in the analysis while the EFTofLSS model remains biased but this bias does not seem to increase significantly with the inclusion of smaller scales. This is likely due to the fact that TNS has one free parameter to describe the damping thus it cannot capture the different damping of $P_0$ and $P_2$. At $k_{\rm max} = 0.186h/{\rm Mpc}$ the fractional $2\sigma$ marginalised error on $f$ is found to be $3.6\%$ and $3.1\%$ for TNS and EFTofLSS respectively. 
\newline
\newline
Secondly, we immediately note from Fig.~\ref{dm1} that there are significant degeneracies between $f$ and the model nuisance parameters. This is not a new result for the TNS model \cite{Zheng:2016xvo,Bose:2017myh}. What we find here though is that this degeneracy doesn't weaken as we go to smaller scales. For the EFTofLSS we can see degeneracy between the sound speed parameters and $f$ as well. This degeneracy seems to weaken as we push to non-linear scales, especially between $c_{s,0}^2$, $c_{s,2}^2$ and $f$. The degeneracy between $c_{s,0}^2$ and $c_{s,2}^2$ is also very clear\footnote{Note that $c_{s,4}^2$ only comes with powers of $\mu$ greater or equal to 4 and so in the monopole and quadrupole it does not significantly contribute.}. 
\newline
\newline
These results can almost be directly compared to those from \cite{delaBella:2017qjy} who also use simulations using the Multidark cosmology in their comparisons. They consider $z=0$ and find that $P_0$ and $P_2$ match the simulation measurements within $2\%$ and $25\%$ respectively up to $k=0.400h/{\rm Mpc}$. These results are significantly higher than what we find here, when we consider bias of cosmological parameters as our criterion for $k_{\rm max}$. Our $k_{\rm max}$ is more comparable to \cite{Lewandowski:2015ziq} who determine a reliable scale of $k=0.130h/{\rm Mpc}$ at $z = 0.56$, using a percent level deviation criterion.   

\subsection{Halos}

\begin{figure}[H]
\centering
  \includegraphics[width=12cm,height=8cm]{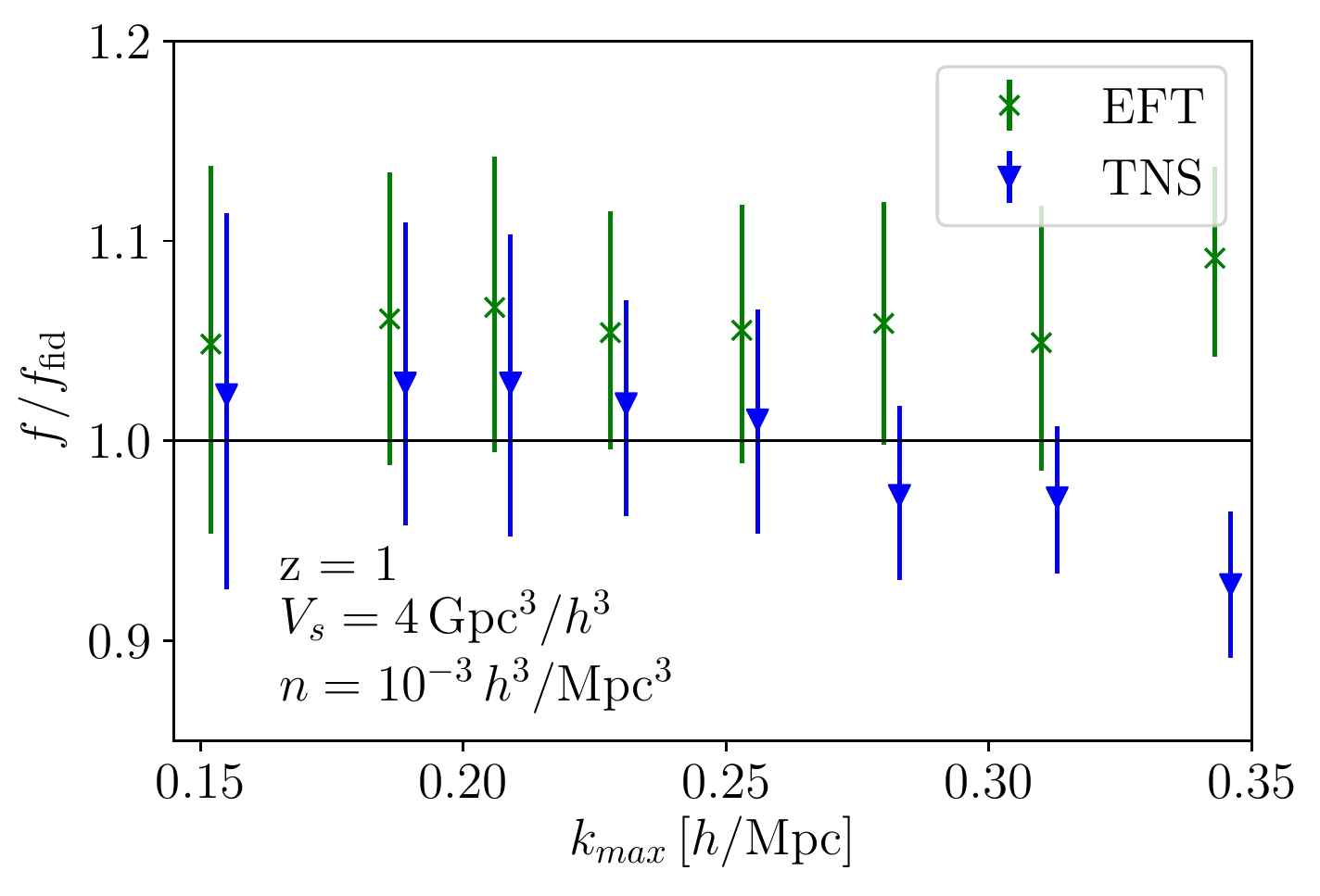}  
  \caption[CONVERGENCE]{ Redshift space halo results at $z=1$ with $V = 4 {\rm Gpc}^3/h^3$ taken as the bin volume and using a halo number density of $n=10^{-3} h^3/{\rm Mpc}^3$ both in selecting the halo catalog and in the analytic covariance matrix used in the analyses. We show the mean value of $f/f_{\rm fiducial}$ as a function of $k_{\rm max}$ using the TNS (blue triangles) and EFTofLSS (green crosses) models with the marginalised $2\sigma$ error bars. Only $P_0$ and $P_2$ were used in the analyses.}
\label{halo1}
\end{figure}

\begin{figure}[H]
\centering
  \includegraphics[width=15cm,height=15cm]{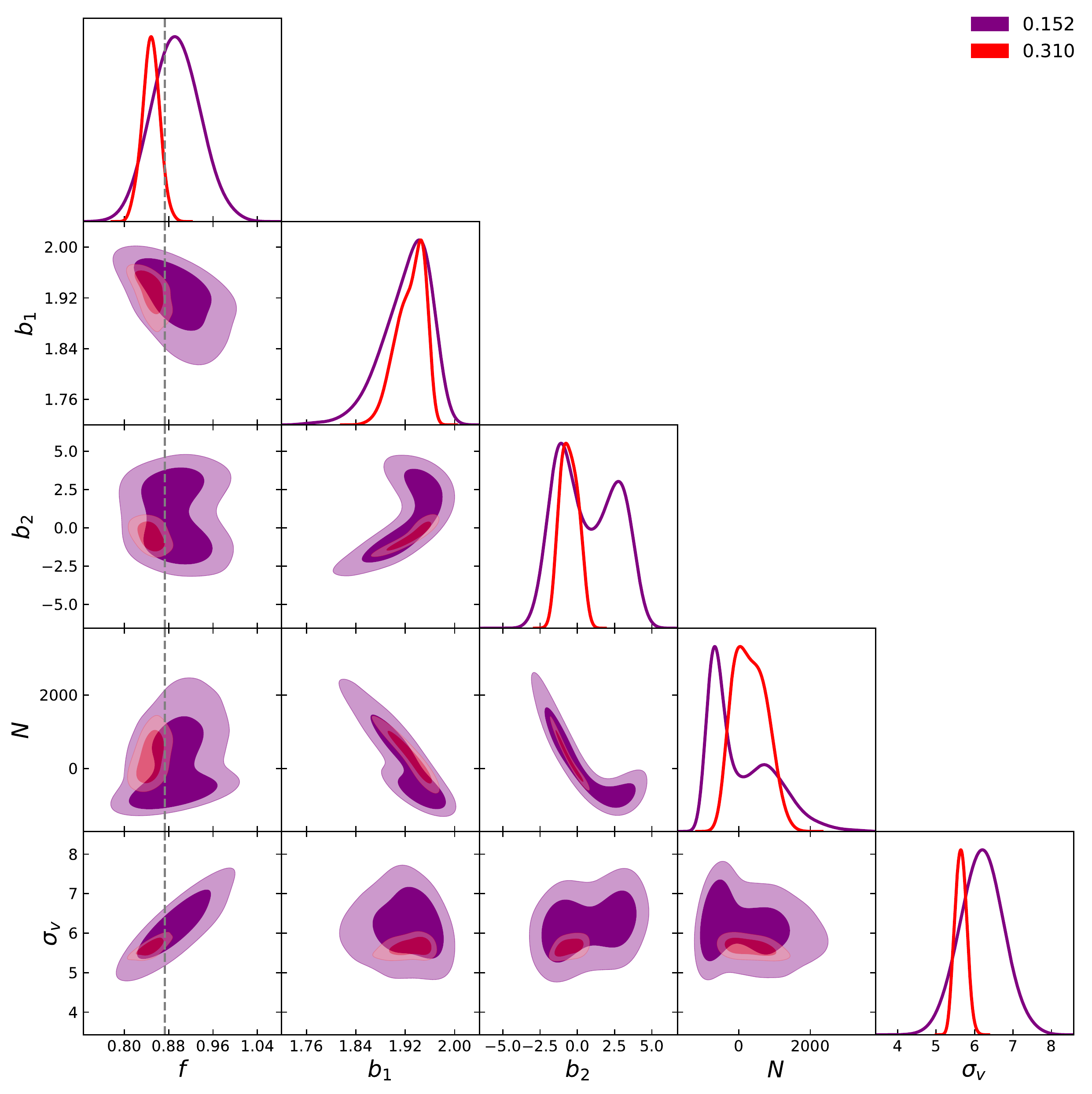}  
  \caption[CONVERGENCE]{Redshift space halo results at $z=1$ with $V = 4 {\rm Gpc}^3/h^3$ taken as the bin volume and using a halo number density of $n=10^{-3} h^3/{\rm Mpc}^3$ both in selecting the halo catalog and in the analytic covariance matrix used in the analyses. The $1\sigma$ and $2\sigma$ confidence contours for the TNS model for $k_{\rm max} = 0.152 h/{\rm Mpc}$ (purple) and $k_{\rm max} = 0.310 h/{\rm Mpc}$ (red). The fiducial value of $f$ is denoted by a dashed line and only $P_0$ and $P_2$ were used in the analyses. }
\label{halo3}
\end{figure}

\begin{figure}[H]
\centering
 \includegraphics[width=18cm,height=18cm]{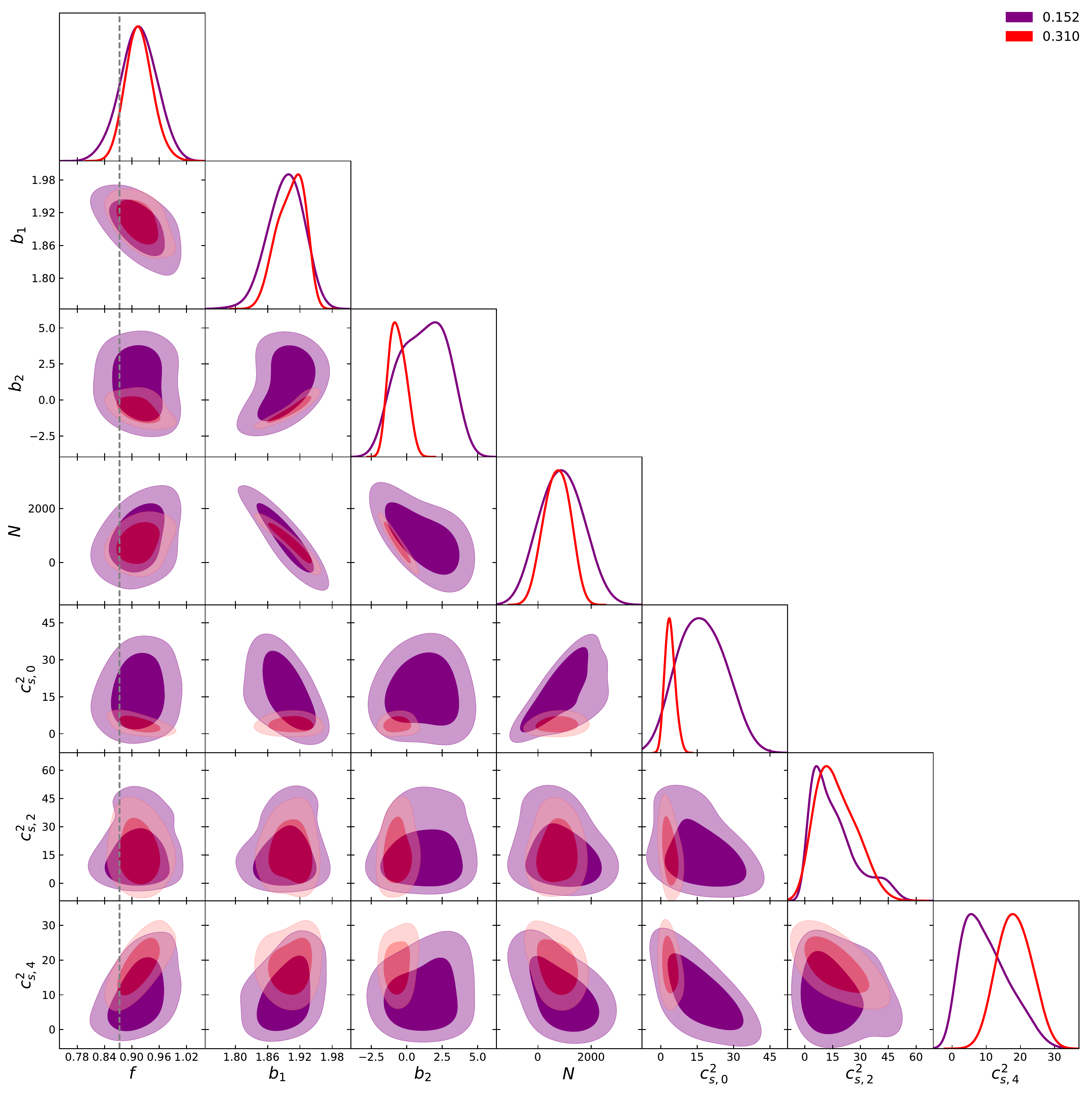}  
  \caption[CONVERGENCE]{Same as Fig.~\ref{halo3} but for the EFTofLSS model.}
\label{halo4}
\end{figure}

Here we investigate the halo multipoles. Fig.~\ref{halo1} shows the $2\sigma$ marginalised constraints on $f$ from the MCMC analyses as a function of $k_{\rm max}$ for both TNS and EFTofLSS models. We see that again, under a $2\sigma$ criterion, the models seem to do equally well, and become biased at around $k_{\rm max} = 0.310h/{\rm Mpc}$, which is significantly larger than the dark matter case. This suggests the bias model provides much added freedom to the models. Further, for TNS, the velocity dispersion of halos is found to be significantly less indicating that different damping of the monopole and quadrupole is not as necessary as in the dark matter case. At $k_{\rm max} = 0.310/{\rm Mpc}$ the fractional $2\sigma$ marginalised error on $f$ is found to be $3.8\%$ and $6.0\%$ for TNS and EFTofLSS respectively. 
\newline
\newline
In Fig.~\ref{halo3} and Fig.~\ref{halo4} we plot the two-dimensional posterior distributions for TNS and EFTofLSS respectively at two different $k_{\rm max}$; $0.152h/{\rm Mpc}$ and $0.310h/{\rm Mpc}$. In the TNS model the degeneracy of $f$ with $\sigma_v$ persists but is significantly reduced by the inclusion of non-linear information. Further, degeneracies of $f$ with the bias parameters is also reduced with the inclusion of smaller scales. This results in a significant improvement of constraints, with the fractional $2\sigma$ marginalised error on $f$ going from $6.9\%$ to $3.8\%$. In the EFTofLSS case we find a marginal improvement on the fractional  error on $f$ from $6.9\%$ to $6.0\%$. This is unsurprising due to the larger number of nuisance parameters we need to marginalise over. 

\newpage

\subsection{Analysis at \texorpdfstring{$z=0.5$}{}}
\begin{figure}[H]
\centering
  \includegraphics[width=12cm,height=8cm]{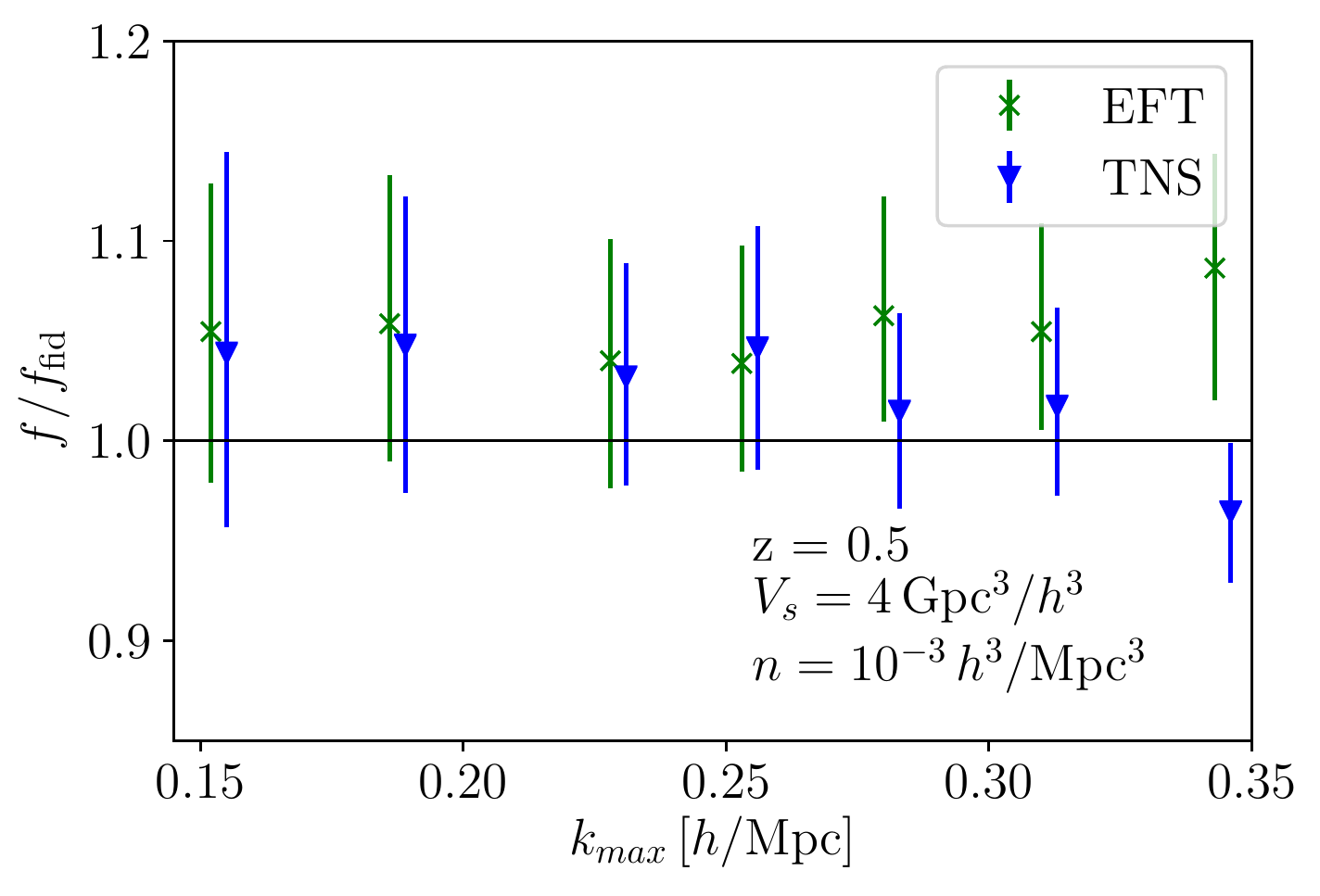}  
  \caption[CONVERGENCE]{Same as Fig.~\ref{halo1} but for $z=0.5$.}
\label{halo5}
\end{figure}

\begin{figure}[H]
\centering
  \includegraphics[width=15cm,height=15cm]{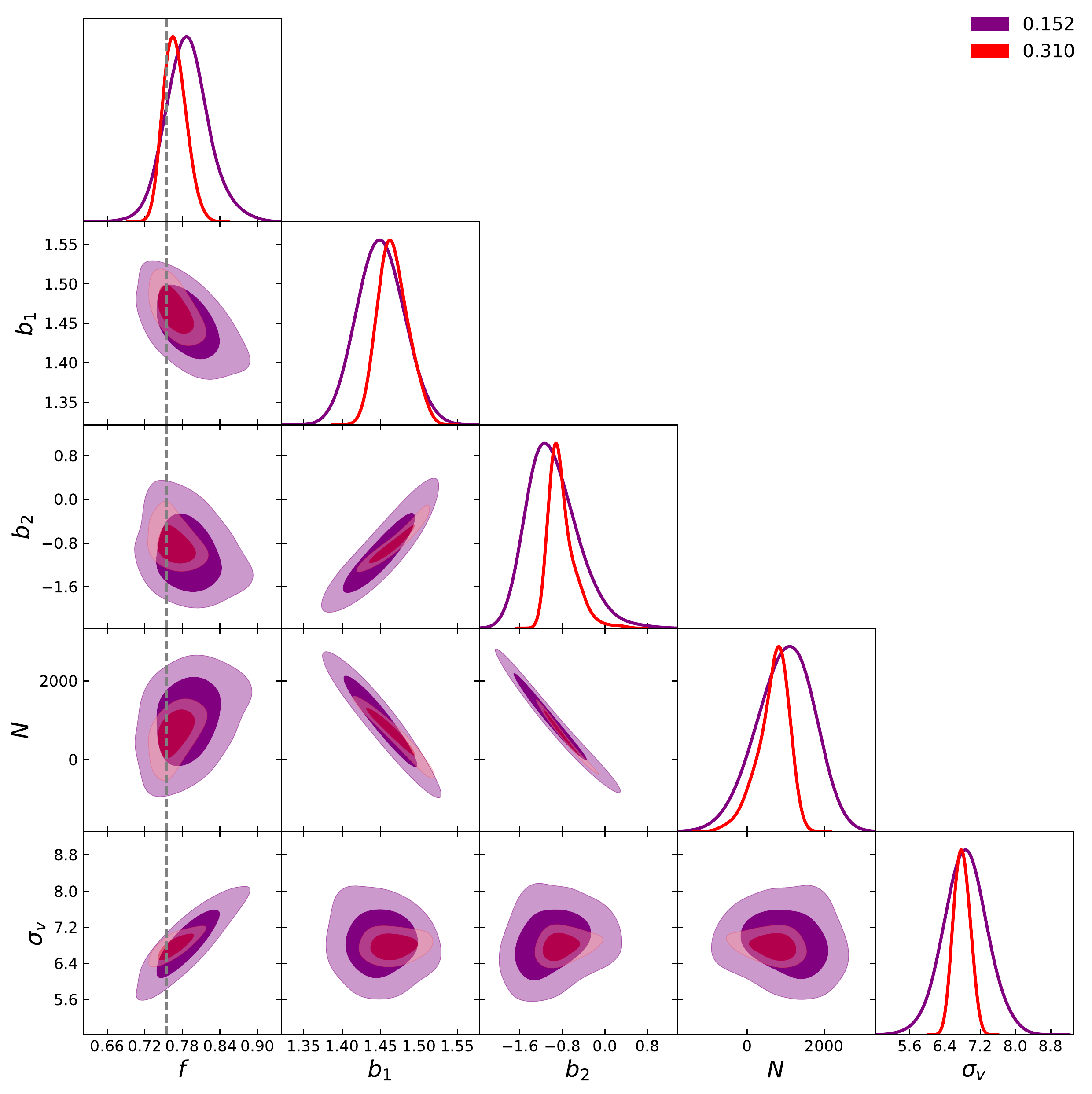}  
  \caption[CONVERGENCE]{ Same as Fig.~\ref{halo3} but at $z=0.5$.}
\label{halo6}
\end{figure}

\begin{figure}[H]
\centering
 \includegraphics[width=18cm,height=18cm]{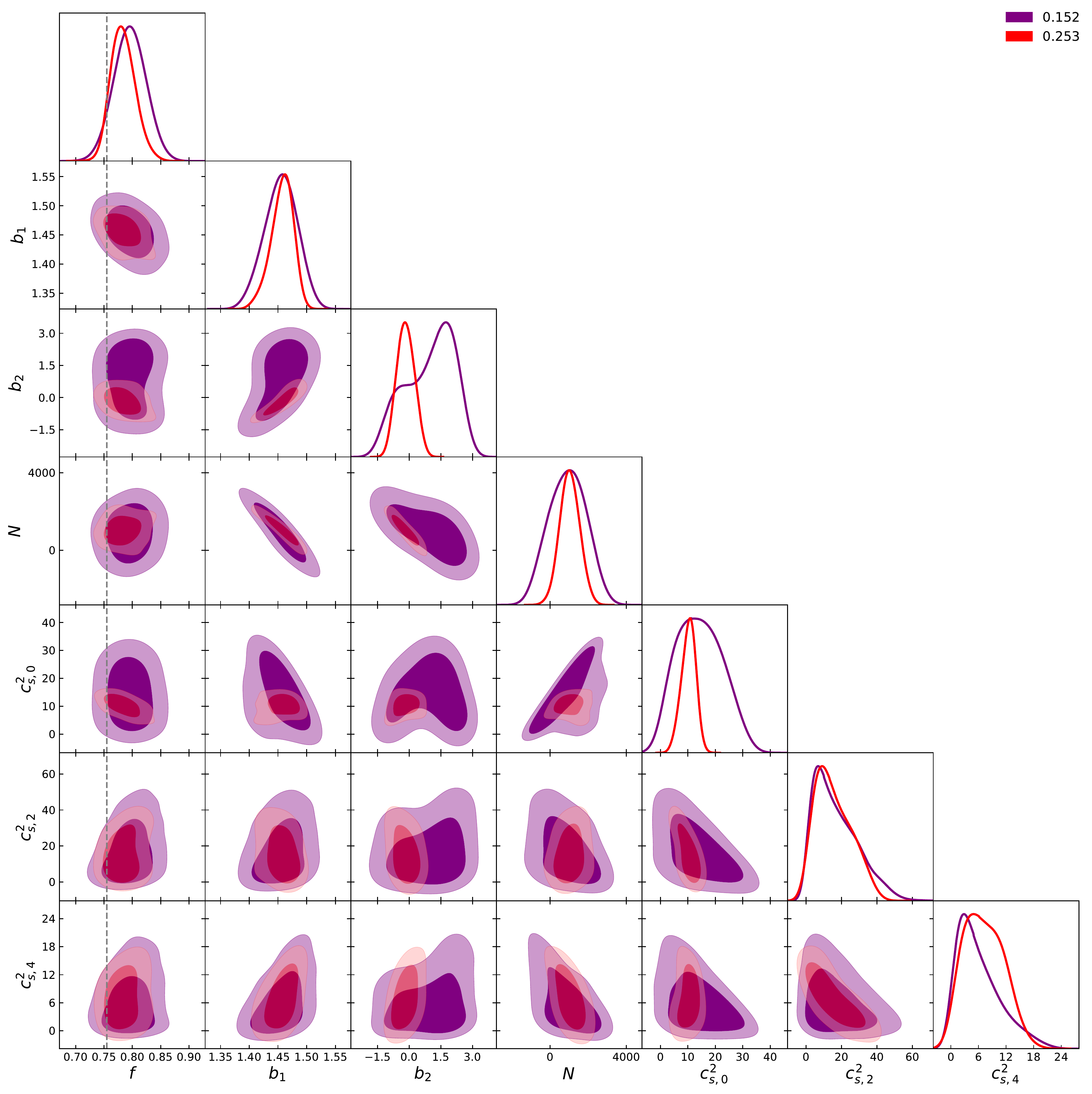}  
  \caption[CONVERGENCE]{Same as Fig.~\ref{halo6} but for the EFTofLSS model.}
\label{halo7}
\end{figure}

Here we consider the same survey volume and number density as in the previous subsection but repeat the analysis for $z=0.5$. Fig.~\ref{halo5} again shows the mean value of $f$ obtained from MCMC analyses using $V=4{\rm Gpc}^3/h^3$ and $n=10^{-3} h^3/{\rm Mpc}^3$ as a function of $k_{\rm max}$. Here we see that again the TNS model does very well in capturing the non-linear RSD halo multipoles without biasing estimates of $f$. On the other hand, the EFTofLSS model becomes slightly biased at $k_{\rm max}> 0.253h/{\rm Mpc}$. Using the same $2\sigma$ criterion, we have $k_{\rm max} = 0.310h/{\rm Mpc}$ for the TNS and $k_{\rm max}=0.253h/{\rm Mpc}$ for the EFTofLSS models. At these $k_{\rm max}$ we find the fractional $2\sigma$ marginalised error on $f$ to be $4.6\%$ and $5.2\%$ for TNS and EFTofLSS respectively. We find at $k_{\rm max}=0.253h/{\rm Mpc}$ the TNS gives a $5.8\%$ error which is in fact worse than EFTofLSS despite the smaller parameter space. This is consistent with what we find in \cite{Bose:2019psj}, but in that work we did not provide a robust check for the 'true' $k_{\rm max}$. 
\newline
\newline
We again plot the two-dimensional posterior distributions  in Fig.~\ref{halo6} and Fig.~\ref{halo7} for TNS and EFTofLSS respectively at two different $k_{\rm max}$; $0.152h/{\rm Mpc}$ and $0.310h/{\rm Mpc}$ for TNS and $0.152h/{\rm Mpc}$ and $0.253h/{\rm Mpc}$ for EFTofLSS. We find that the parameter degeneracies do not change significantly between $z=1$ and $z=0.5$ for both models. We also find that at $z=0.5$ we get similar gains in the constraints on $f$ by going to smaller scales. This improves its fractional $2\sigma$ marginalised error on $f$ from $9.1\%$ to $4.6\%$ while the EFTofLSS gains less due to its lower $k_{\rm max}$, going from $7.2\%$ to $5.2\%$. 
\newline
\newline
So far the TNS model seems to outperform the EFTofLSS model considered here, but this may be because we are not fully utilising all the EFTofLSS free parameters. To better test the capabilities of this model we will next include $P_4$ in our analyses.

\subsection{Impact of hexadecapole}
\begin{figure}[H]
\centering
  \includegraphics[width=12cm,height=8cm]{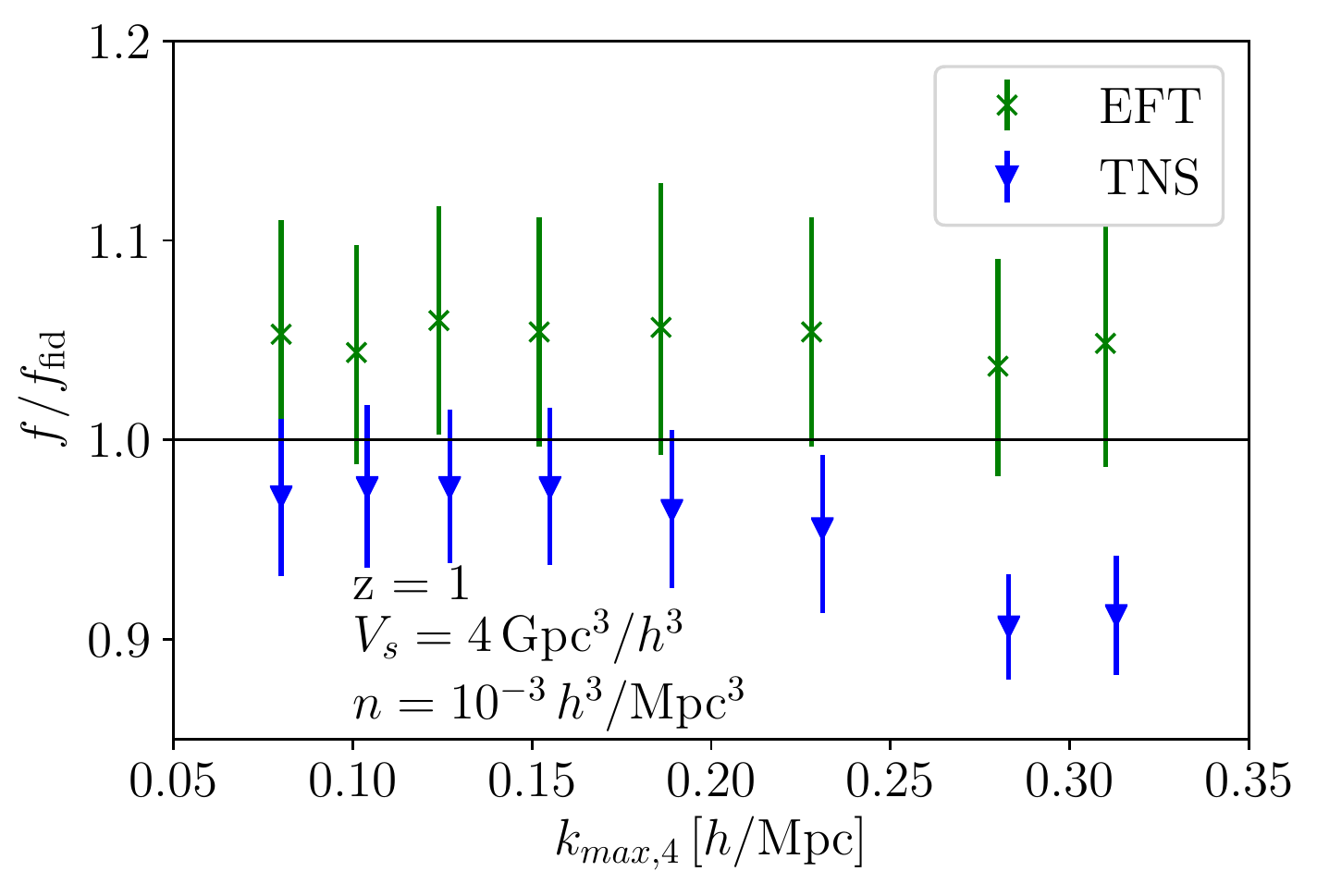}  
  \caption[CONVERGENCE]{ Redshift space halo results at $z=1$ with $V = 4 {\rm Gpc}^3/h^3$ taken as the bin volume and using a halo number density of $n=10^{-3} h^3/{\rm Mpc}^3$ both in selecting the halo catalog and in the analytic covariance matrix used in the analyses. Here we fix $k_{\rm max} = 0.310 h/{\rm Mpc}$ and include $P_4$ and the relevant covariance in the likelihood only up to $k_{\rm max,4}$. We show the mean value of $f/f_{\rm fiducial}$ as a function of $k_{\rm max,4}$ using the TNS (blue triangles) and EFTofLSS (green crosses) models with the marginalised $2\sigma$ error bars.} 
  
\label{halo8}
\end{figure}

\begin{figure}[H]
\centering
  \includegraphics[width=12cm,height=8cm]{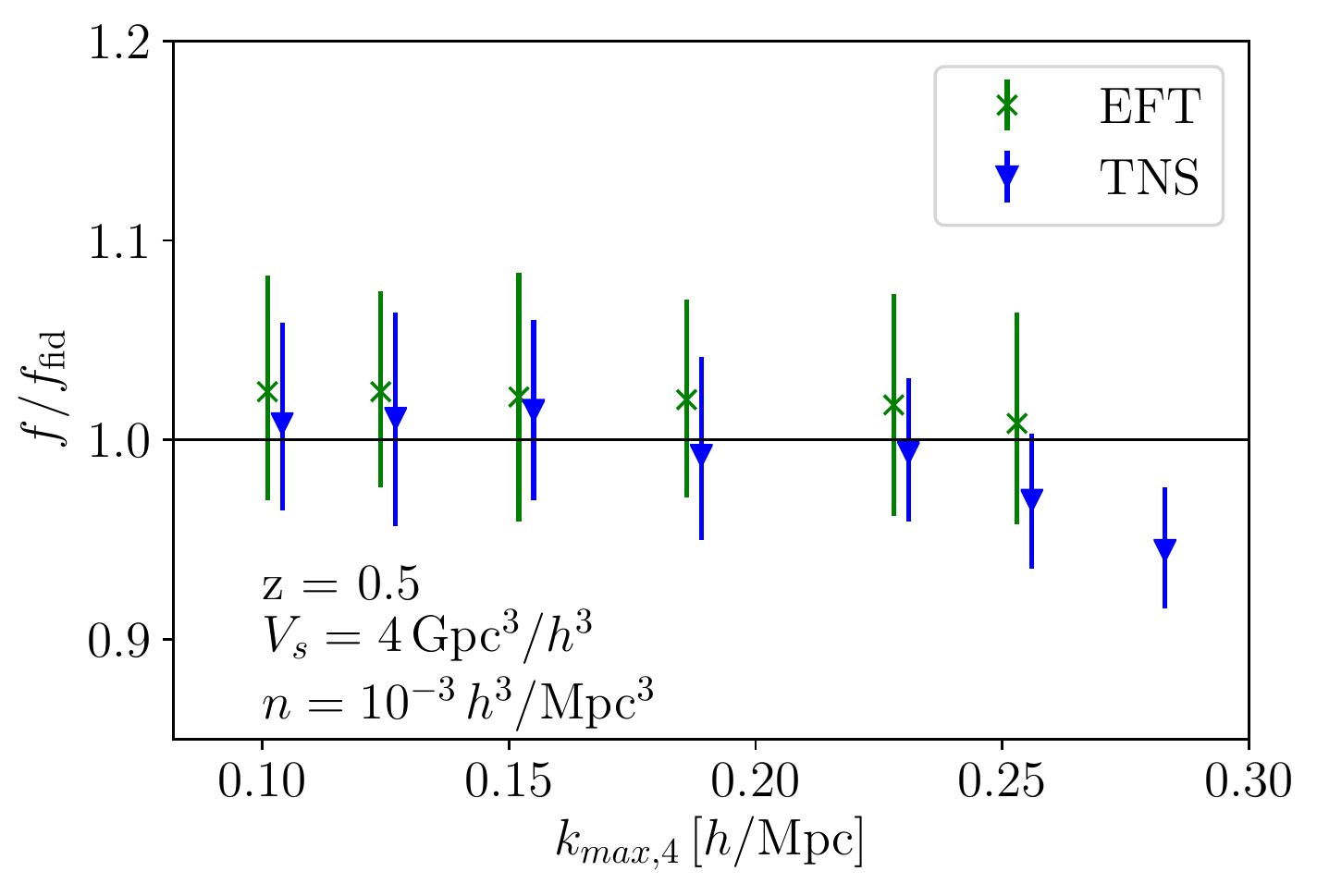}  
  \caption[CONVERGENCE]{Same as Fig.~\ref{halo8} but at $z=0.5$.}
\label{halo9}
\end{figure}

\begin{figure}[H]
\centering
  \includegraphics[width=15cm,height=15cm]{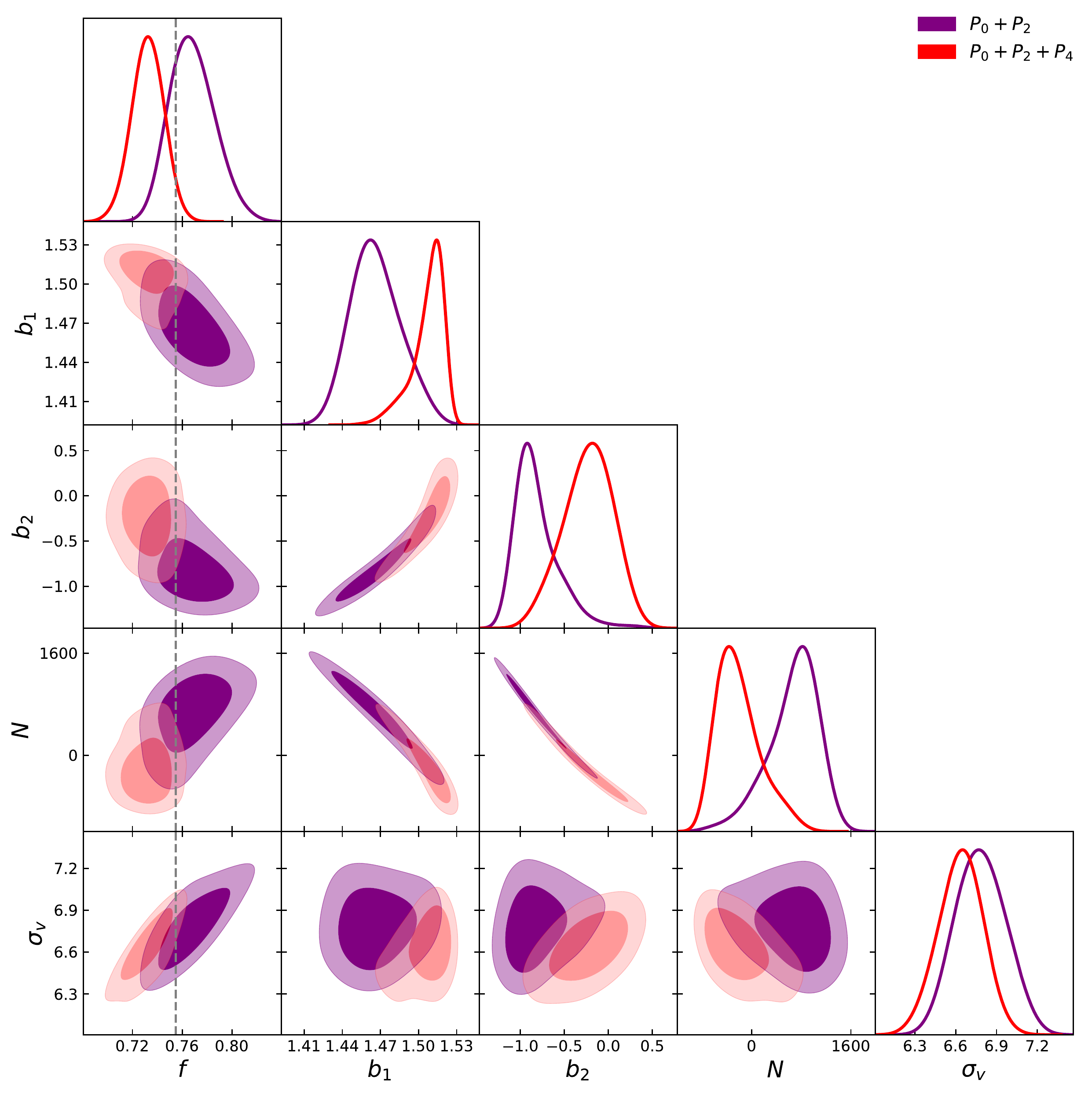}  
  \caption[CONVERGENCE]{Redshift space halo results at $z=0.5$ with $V = 4 {\rm Gpc}^3/h^3$ taken as the bin volume and using a halo number density of $n=10^{-3} h^3/{\rm Mpc}^3$ both in selecting the halo catalog and in the analytic covariance matrix used in the analyses. The $1\sigma$ and $2\sigma$ confidence contours for the TNS model for $k_{\rm max} = 0.310h/{\rm Mpc}$ and $k_{\rm max,4} = 0.253 h/{\rm Mpc}$ without $P_4$ (purple) and with $P_4$(red). The fiducial value of $f$ is denoted by a dashed line.}
\label{halo10}
\end{figure}

\begin{figure}[H]
\centering
  \includegraphics[width=18cm,height=18cm]{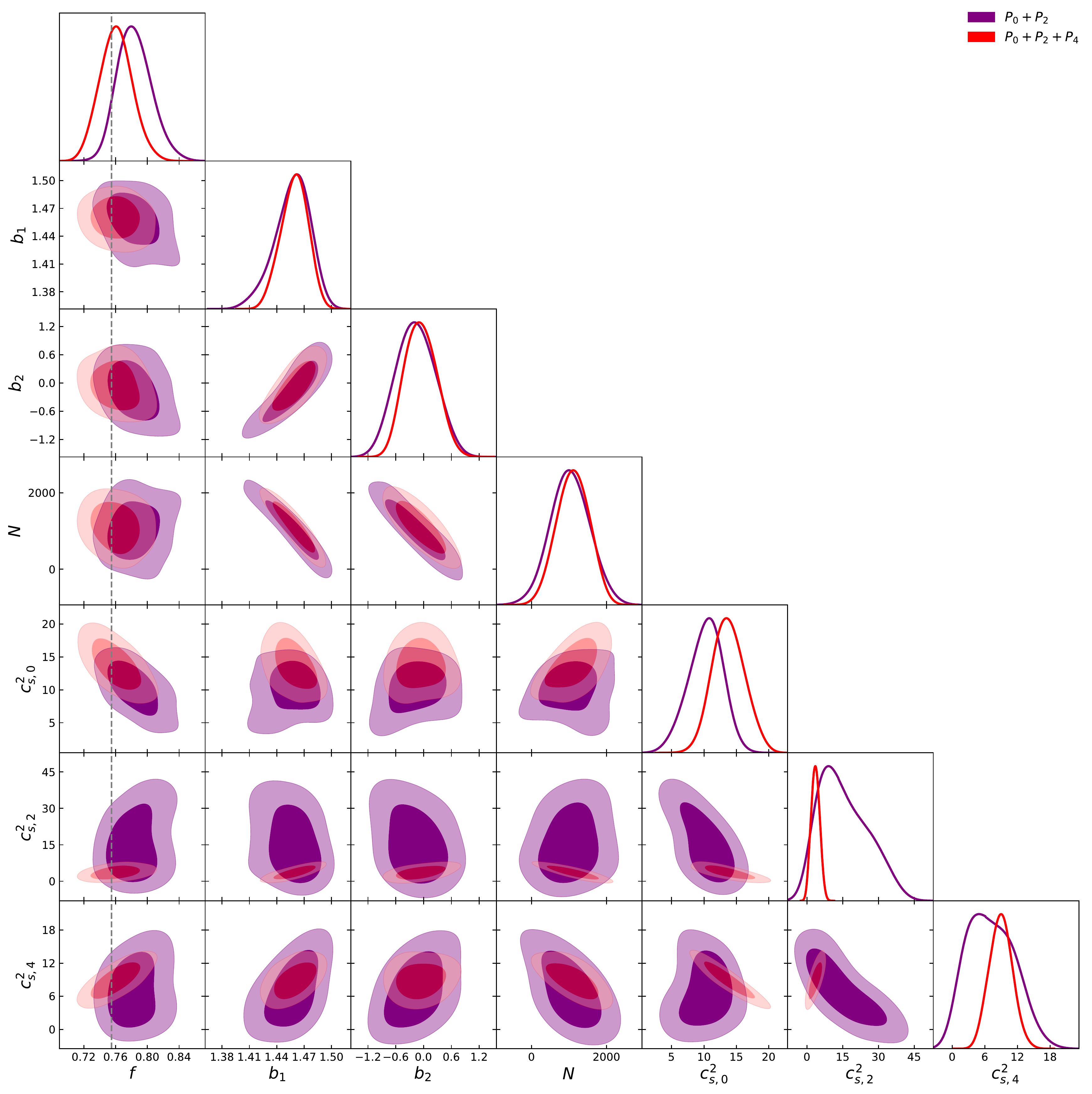}  
  \caption[CONVERGENCE]{Same as Fig.~\ref{halo10} but for the EFTofLSS model. Here $k_{\rm max} = k_{\rm max,4} = 0.253 h/{\rm Mpc}$.}
\label{halo11}
\end{figure}

In this subsection we include $P_4$ in the analysis. As mentioned previously, this would severely restrict the range of scales allowed in the analysis since in general modelling of $P_4$ has been found to be poor for the TNS model \cite{Beutler:2016arn,Markovic:2019sva}, and taking it to too high a $k_{\rm max}$ can result in a biased estimate for $f$. We thus proceed by fixing $k_{\rm max}$ for $P_0$ and $P_2$ as determined in the previous section (see Fig.~\ref{halo1} and Fig.~\ref{halo5}). Then, we include $P_4$ up to a new $k_{\rm max,4}$ testing for bias of $f$ at the $2\sigma$ level as before. Note that above $k_{\rm max,4}$ we only include terms with $\ell =0,2$ in the likelihood given in Eq.~\ref{covarianceeqn}. We do this for both $z=0.5$ and $z=1$. 
\newline
\newline
Fig.~\ref{halo8} and Fig.~\ref{halo9} shows the marginalised constraints as a function of $k_{\rm max,4}$ at $z=1$ and $z=0.5$ respectively. At $z=1$ we see that the TNS becomes slowly biased, with a determined $k_{\rm max,4} = 0.186h/{\rm Mpc}$ while the EFTofLSS remains unbiased up to $k_{\rm kmax,4} = 0.310h/{\rm Mpc}$. A similar trend is seen at $z=0.5$, with the EFTofLSS remaining unbiased all the way up to $k_{\rm max,4}= k_{\rm max} = 0.253h/{\rm Mpc}$  while the TNS becomes biased before $k_{\rm max,4}$ reaches $k_{\rm max} = 0.310h/{\rm Mpc}$, which at the $2\sigma$ level is determined to be $k_{\rm max,4} = 0.253h/{\rm Mpc}$.  This is to be expected as the EFTofLSS allows for individual non-linear damping over the 3 multipoles using all 3 $c_{s,i}^2$ while the TNS only utilises a single free parameter for all 3 multipoles. 
\newline
\newline
We find that at the determined $k_{\rm max,4}$, the $2\sigma$ marginalised fractional errors on $f$ for the TNS model to be  $4.0~(3.8)\%$ and $3.4~(4.6)\%$ while the EFTofLSS model's are $5.7~(6.0)\%$ and $5.1~(5.2)\%$ at $z=1$ and $z=0.5$ respectively, where in brackets we indicate the fractional errors determined only using $P_0$ and $P_2$. This indicates that at $z=1$, given the hexadecapole's weak signal and our errors, it doesn't offer significant additional information on the growth of structure. At $z=0.5$ it becomes more important and improves constraints significantly for the TNS model whereas for the EFTofLSS it again doesn't add to the constraints. To gain insight into this we look at the two-dimensional posterior distributions. This is shown in Fig.~\ref{halo10} and Fig.~\ref{halo11} for the TNS and EFTofLSS respectively. In the TNS case, $P_4$ both affects the degeneracy and the constraints on $\sigma_v$, which has a strong degeneracy with $f$. This results in the noticeable improvement of the marginalised constraints on $f$. In the EFTofLSS case, although $P_4$ improves the constraints on the value of $c_{s,2}^2$ and $c_{s,4}^2$, these lack a strong degeneracy with $f$ (unlike $c_{s,0}^2$) and so constraints on $f$ are not improved. Further, by including $P_4$ we move the best fit sound speed parameters, $c_{s,i}^2$, to larger values which reduces the impact of the positivity priors we impose on these parameters. This can explain why at $z=0.5$ including $P_4$ does not improve the constraint on $f$ in the EFTofLSS case. 
\newline
\newline
We also comment on the effect of $P_4$ at $z=1$ in the TNS case. Given the small $k_{\rm max,4}=0.186h/{\rm Mpc}$, we don't expect $P_4$ to make a noticeable impact. By inspection of the two-dimensional posteriors between the $P_0+P_2$ and the $P_0+P_2+P_4$ analyses (not shown in the paper), we find the parameter degeneracies are not impacted by $P_4$ and it only serves to sharpen most of the marginalised distributions, with the exception of $f$ which broadens slightly. Since the constraint on $f$ does not degrade significantly we do not investigate this matter further. 
\newline
\newline
Of course all the results so far rely on the errors and scatter of the data. The next three subsections are dedicated to investigating these issues. We first look at the effect of smaller errors. 

\subsection{Analysis with \texorpdfstring{$V = 15{\rm Gpc}^3/h^3$}{}}

\begin{figure}[H]
\centering
  \includegraphics[width=12cm,height=8cm]{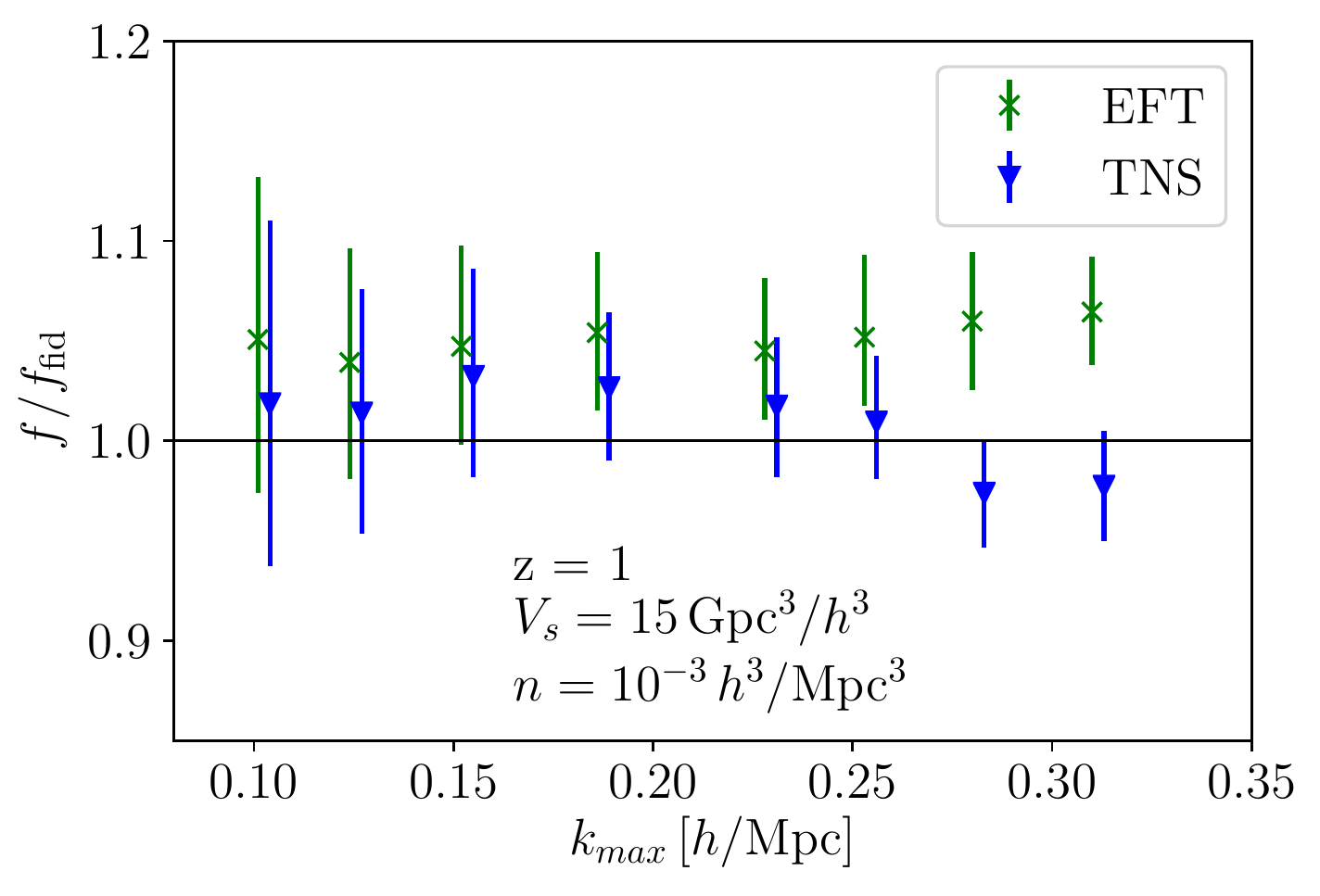}  
  \caption[CONVERGENCE]{ Redshift space halo results at $z=1$ with $V = 15 {\rm Gpc}^3/h^3$ taken as the total survey volume and using a halo number density of $n=10^{-3} h^3/{\rm Mpc}^3$ both in selecting the halo catalog and in the analytic covariance matrix used in the analyses. We show the mean value of $f/f_{\rm fiducial}$ as a function of $k_{\rm max}$ using the TNS (blue triangles) and EFTofLSS (green crosses) models with the marginalised $2\sigma$ error bars. Only $P_0$ and $P_2$ were used in the analyses.}
\label{halo12}
\end{figure}

In this section we again concentrate on $z=1$ and the number density $n=10^{-3} h^3/{\rm Mpc}$ but consider a much larger survey volume, $V=15 {\rm Gpc}^3/h^3$ which will give a better representation of the errors when using many redshift bins and of the total power of upcoming surveys. 
\newline
\newline
Again, we show the $2\sigma$ marginalised constraints on $f$ from the MCMC analyses as a function of $k_{\rm max}$ for both TNS and EFTofLSS models in Fig.~\ref{halo12}. Immediately we note the EFTofLSS model becomes biased very quickly with a new $k_{\rm max}=0.152h/{\rm Mpc}$ which is to be expected from Fig.~\ref{halo1} where all the mean values from the MCMC analyses lie significantly above the fiducial. On the other hand the TNS model seems robust against the reduced error bars and barely maintains its original $k_{\rm max}=0.310h/{\rm Mpc}$. We find the $2\sigma$ marginalised fractional errors on $f$ at these $k_{\rm max}$ are $2.8~(3.8)\%$ and $4.7~(6.0)\%$ for TNS and EFTofLSS respectively where the bracketed values are from the analysis using $V=4{\rm Gpc}^3/h^3$. Naturally the TNS constraints are improved significantly maintaining the same $k_{\rm max}$, but we also note that the EFTofLSS results also improve despite the far lower $k_{\rm max}$. If we maintain the same $k_{\rm max} = 0.310h/{\rm Mpc}$ the EFTofLSS produces a $2.5\%$ fractional error, but its mean value of $f$ is biased by over $4\sigma$. Similarly, if we take the TNS model to $k_{\rm max}=0.152h/{\rm Mpc}$ we find a fractional error of $4.9\%$ which is comparable to that of the EFTofLSS model. 

\subsection{Analysis with \texorpdfstring{$n=10^{-4} h^3/{\rm Mpc}^3$}{}}

\begin{figure}[H]
\centering
  \includegraphics[width=12cm,height=8cm]{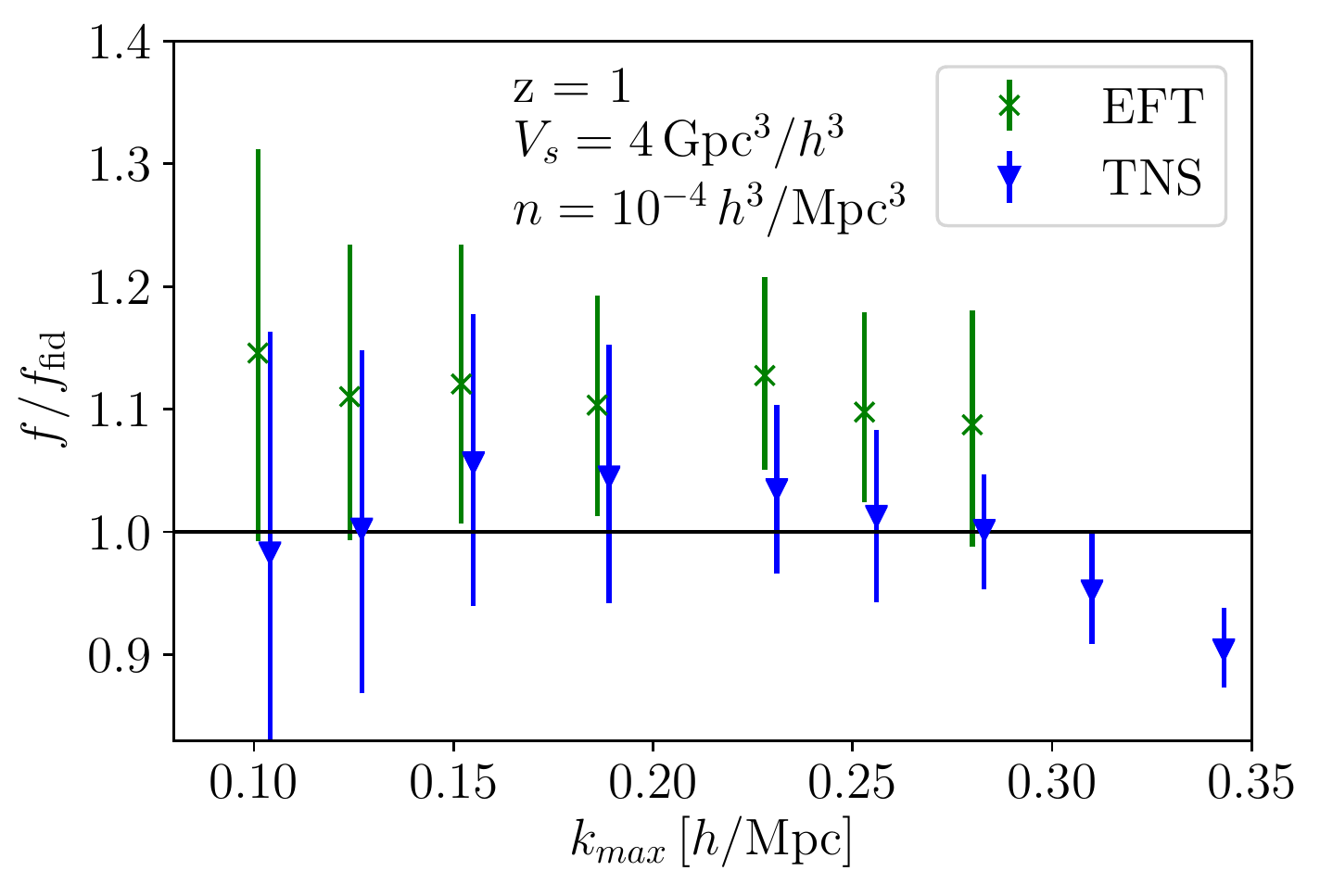}  
  \caption[CONVERGENCE]{ Redshift space halo results at $z=1$ with $V = 4 {\rm Gpc}^3/h^3$ taken as the total survey volume and using a halo number density of $n=10^{-4} h^3/{\rm Mpc}^3$ both in selecting the halo catalog and in the analytic covariance matrix used in the analyses. We show the mean value of $f/f_{\rm fiducial}$ as a function of $k_{\rm max}$ using the TNS (blue triangles) and EFTofLSS (green crosses) models with the marginalised $2\sigma$ error bars. Only $P_0$ and $P_2$ were used in the analyses.}
\label{halo13}
\end{figure}

Here we investigate the impact of taking a catalog of more massive halos which translates to a lower number density. On the theoretical side, in the analytic covariance we use $b_1=2.95$ and $n=10^{-4}h^3/{\rm Mpc}^3$ while keeping $V=4{\rm Gpc}^3/h^3$ while the halo catalog measured from simulations makes a number density cut of $n=10^{-4}h^3/{\rm Mpc}^3$. This naturally introduces larger errors through the analytic covariance prescription used in Eq.~\ref{covarianceeqn} as well as larger scatter in the data due to a lower number of halos. These halos are more massive and so also more biased allowing a test for the flexibility of the bias model, as well as its compatibility with each RSD model.   
\newline
\newline
Once again, Fig.~\ref{halo13} shows the $2\sigma$ marginalised constraints on $f$ from the MCMC analyses as a function of $k_{\rm max}$ for both models. Similar to Fig.~\ref{halo12}, where we considered a larger survey volume, we find the TNS model remains robust achieving  $k_{\rm max} = 0.310h/{\rm Mpc}$. The EFTofLSS model on the other hand becomes biased much earlier at $k_{\rm max} = 0.124h/{\rm Mpc}$, but this bias seems to be within $3-4\sigma$ over a large range of $k_{\rm max}$. Again, the fractional errors at the determined $k_{\rm max}$ are $4.8 ~(3.8)\%$ and $10.1~(6.0)\% $  for the TNS and EFTofLSS model respectively, where the bracketed value is that obtained from the $n=10^{-3}h^3/{\rm Mpc}^3$ analysis. 
\newline
\newline
To investigate why the EFTofLSS fails at such a small $k_{\rm max}$ we provide a test in Appendix~\ref{app:bias} where we compare the best fit value of $b_1$ over the full range of $k_{\rm max}$ to that measured from the simulations. This gives a good indication of what scales the bias model works well at. In particular, we refer the reader to Fig.~\ref{bias3}, which supports a failure in the bias model when considering highly biased tracers as seen in Fig.~\ref{halo13} for the EFTofLSS.  Further, in \cite{Lewandowski:2015ziq,Fujita:2016dne} which consider the full biased tracer model for EFTofLSS \cite{Perko:2016puo}, they consider terms which scale with halo mass. These typically go as $~k^4$ or are partially degenerate with the counter terms considered here. Since we observe a biased recovery of $f$ in the EFTofLSS-like model we consider at $k=0.124h/{\rm Mpc}$ we do not expect these terms to rescue this model. Further, in Appendix~\ref{app:bias} we also find that the Roy and McDonald model seems to break down at the same scales for both TNS and EFTofLSS model and so additional bias terms may be needed in both of these RSD models.

\subsection{Analysis using fewer realisations} 
\begin{figure}[H]
\centering
  \includegraphics[width=12cm,height=8cm]{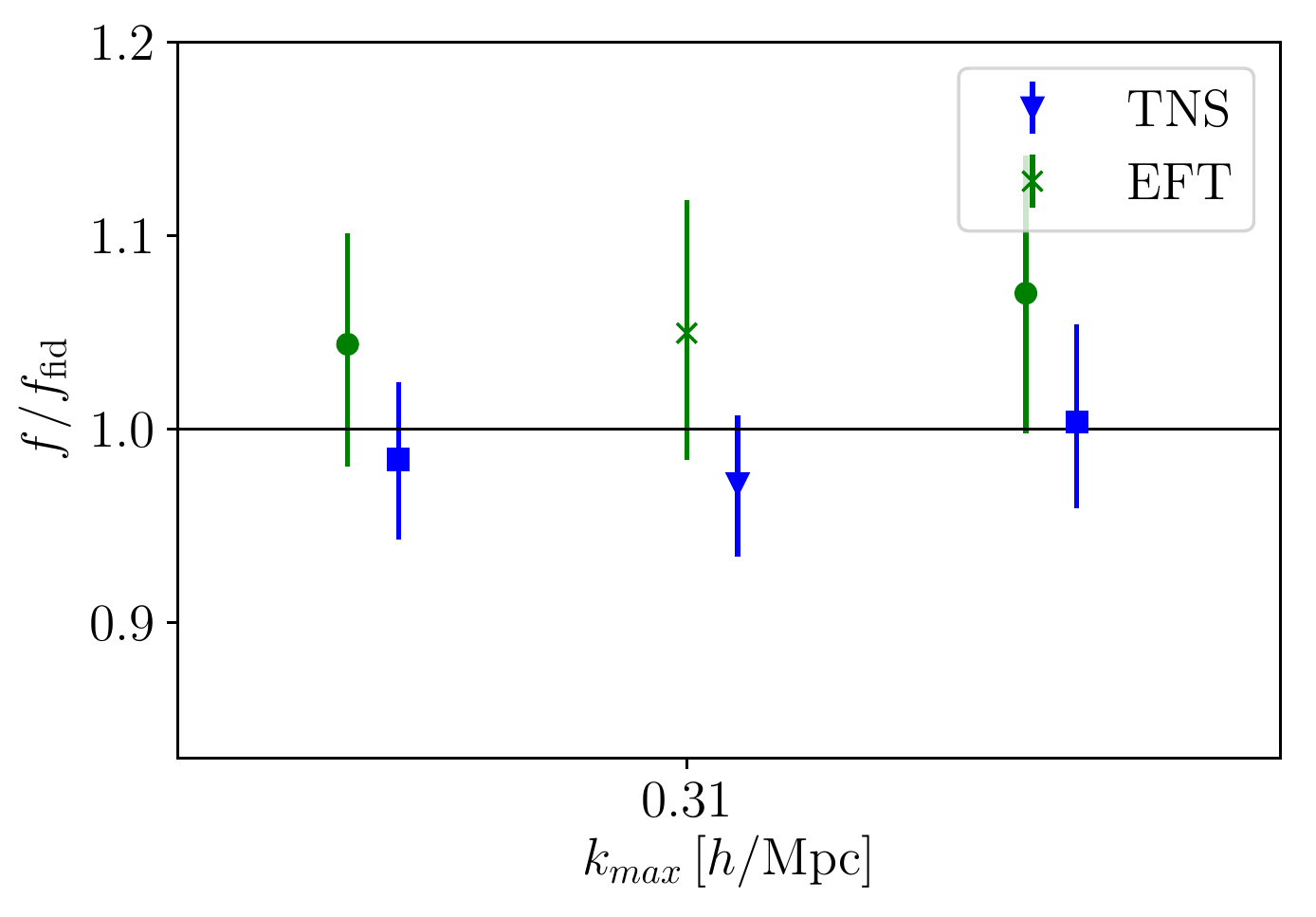}  
  \caption[CONVERGENCE]{Same as Fig.~\ref{halo1} but only at $k_{\rm max}=0.310h/{\rm Mpc}$ (all points except the green cross have been offset for better visualisation). The cross and triangle show the same results as Fig.~\ref{halo1} while squares and circles are the results when we only use an average of 4 realisations rather than 35 for the TNS and EFTofLSS models respectively. Only $P_0$ and $P_2$ were used in the analyses.}
\label{halo14}
\end{figure}

Finally, as mentioned in Sec.~\ref{sims}, our simulation measurements are the average of $35$ realisations and so represent an ideal measurement which are not truly representative of a real observation which will come with scatter which is associated with the errors we've attached. Our goal was to test for bias in the models when modelling non-linearity and so we wanted to use highly converged data. In reality, scatter in the data may affect the $k_{\rm max}$ and in turn introduce a bias if we are to trust a $k_{\rm max}$ determined from mocks. To investigate this issue we consider 2 sets of 4 realisations taken randomly from the original 35. We repeat the analysis at $z=1$, using $V=4 {\rm Gpc}^3/h^3$ and $n=10^{-3}h^3/{\rm Mpc}^3$ for $k_{\rm max} = 0.310h/{\rm Mpc}$ twice, once each using the average of both these sets. 
\newline
\newline
Fig.~\ref{halo14} shows the results at $k_{\rm max} =0.310h/{\rm Mpc}$ for the 2 sets of 4 realisations (shown as circular dots) for the TNS and EFTofLSS models. The plot indicates that the $k_{\rm max}$ we have determined using the 35 realisations is robust against scatter in the data within the given errors. Further, for both models we find the qualitative shape of the contours does not change appreciably for most of the parameters although central positions and overal sizes do shift within $2\sigma$ of the $35$ realisation average contours for some parameter pairs.

\section{Discussion and Conclusion}\label{summary}
In this work we have extended a number of previous analyses \cite{delaBella:2018fdb,Osato:2018ldv,Markovic:2019sva} which attempt to discern which models are most apt to model galaxy clustering for upcoming surveys. In particular, we select the two models identified in \cite{Bose:2019psj} as being contenders; the TNS model with a Lorentzian damping factor and an EFTofLSS based model. The EFTofLSS model is similar to one of the leading models identified in \cite{delaBella:2018fdb}. We extend previous works by completing many MCMC analyses, using high quality PICOLA simulation data, in which we vary the growth rate of structure $f$ as well as all model nuisance parameters (4 for TNS and 6 for EFTofLSS). In particular, we thoroughly test for biased estimation of $f$ by the models when considering quasi non-linear scales. These tests are all conducted within the context of upcoming surveys through our selection of the halo catalogs, the simulation volume and our modelling of the RSD-multipole covariance matrix. Further, we test the robustness of the models by considering a different redshift, the hexadecapole, a different halo catalog, a different survey volume and scatter in the data. All our core results are summarised in Table~\ref{summarytable}. 
\newline
\newline
Overall, we find that the TNS model seems to do better in its constraints on $f$ and range of validity than the EFTofLSS model considered here, despite the EFTofLSS's larger nuisance parameter space. This is not inconsistent with the results of \cite{Bose:2019psj} where a robust test for $k_{\rm max}$ was not performed. In fact at both redshifts we find a much larger $k_{\rm max}$ for the models. Although \cite{Bose:2019psj} find that at $z=1$ both models achieve the same $k_{\rm max} = 0.276h/{\rm Mpc}$, with EFTofLSS giving better marginalised constraints, we find the models push to a higher $k_{\rm max}$ when properly tested for bias, and the inclusion of these smaller scales may give TNS the edge we see here. But, our conclusion here of course comes with a number of caveats. First, we do not vary the Alcock-Paczynski parameters \cite{Alcock:1979mp} nor consider cosmology beyond $f$, and so do not account for degeneracies between nuisance parameters and these. Second, our EFTofLSS model is phenomenological in the sense that we have treated tracer bias in an ad hoc way by bolting on the bias model of \cite{McDonald:2009dh}. Appendix~\ref{app:bias} suggests that this treatment seems to do well for low biased tracers. The proper treatment of bias within the EFTofLSS follows \cite{Perko:2016puo} and includes 4 more nuisance parameters. With so many nuisance parameters, it seems unlikely that one can achieve better performance in terms of constraints as suggested in \cite{delaBella:2018fdb}. But it is still left to be checked if the full biased tracer EFTofLSS model can achieve better constraints than the TNS and we leave that for a future work. Finally, we used an idealised mock data based on dark matter halos and a Gaussian covariance matrix in this work. In order to apply these models to actual observations, we need to take into account the distributions of galaxies within dark matter halos, a survey window function as well as non-Gaussian covariance matrix. These strongly depend on specifications of a specific future survey and we will need to redo the analysis using galaxy mocks designed for the survey.
\newline
\newline
Beyond constraining power, the models each offer their own advantages and disadvantages. 
The TNS model is simpler in terms of number of parameters which makes it computationally preferable especially when performing MCMC analyses with a large parameter space where convergence may become an issue. It is flexible in terms of modelling the small scale fingers-of-god damping. Using the Lorentzian damping, it effectively re-sums an expansion of the damping term in $k^2$. The fact that its constraining power strongly depends on the form of the this damping term indicates that we could improve the model by taking into account the different damping of multipoles. Priors on $\sigma_v$ can also be conceivably achieved through simulations and even observations \cite{Hikage:2011ut}, which would improve the model's constraining abilities. Further, loop corrections in the perturbative part can be added to improve its modelling of the small scales at the considered redshifts. Alternatively, these corrections can be calibrated by simulations \cite{Song:2018afp, Zheng:2018ljz}. The model has also been already extended to general theories of gravity and dark energy \cite{Bose:2016qun,Bose:2017jjx}. 
\newline
\newline
On the other hand, the EFTofLSS model provides a very systematic way of modelling the small scales and also a way of keeping track of theoretical uncertainties.  This will be very important for upcoming surveys where percent level accuracy is needed. It also provides more flexibility in modelling higher order multipoles as seen here, without biasing $f$. Further, priors on the sound speed parameters, $c_{s,i}^2$ can be achieved through multiple redshift measurements and a knowledge of their dependency on redshift \cite{Foreman:2015uva}. There have also been some attempts to extend this model to modified theories of gravity and dark energy \cite{Lewandowski:2016yce,Cusin:2017wjg,Bose:2018orj}.
\newline
\newline
Independent of model comparisons, we find that the TNS with a Lorentzian damping factor as well as the perturbative components modelled within SPT is a very good prescription for galaxy clustering modelling, achieving $k_{\rm max}=0.310h/{\rm Mpc}$ at $z=1$ and $z=0.5$. In previous analyses this particular form of the TNS model was not considered, and rather a Gaussian damping was used with a RegPT \cite{Taruya:2012ut} prescription for the perturbative components \cite{Beutler:2013yhm,Beutler:2016arn} which was found to have significantly worse fits to simulations in \cite{Bose:2019psj} at the redshifts considered here. A feature left to be desired of this model is an ability to model the hexadecapole up to a larger $k_{\rm max}$. Ways to include the hexadecapole in an optimal way is left to a future work. Further, in principle, for a self-consistent joint data analysis of lensing and galaxy clustering, across a wide range of scales, one would need the same input matter power spectrum. Perturbative models for lensing are highly restrictive and so including a non-linear matter spectrum as input for RSD modelling is also something the authors are highly interested in. This would be very relevant for upcoming surveys that perform both lensing and clustering such as Euclid.


\begin{table}[h]
\centering
\caption{{\bf Summary of results:} $2\sigma$ marginalised fractional errors on $f$ from the MCMC analyses with $k_{\rm max} [h/{\rm Mpc}]$ indicated in curved brackets and $k_{\rm max,4}$ indicated with parentheses. Note analyses with $P_4$ assume $k_{\rm max}$ determined from the $P_0 + P_2$ only analysis. }
\begin{tabular}{| c | c | c | c | c | c |}
\hline  
  Multipoles  & $z$ & $n[h^3/{\rm Mpc}^3]$  &    $V[{\rm Gpc}^3/h^3]$  & TNS-based model & EFTofLSS-based model  \\
 \hline
  $P_0+P_2$  & 1 & $10^{-3}$ & 4 & $3.8\%~~(0.310)$ & $6.0\%~~(0.310)$  \\ \hline 
  $P_0+P_2 + P_4$  & 1 & $10^{-3}$ & 4 & $4.0\%~~(0.187)$  & $5.7\%~~(0.310)$  \\ \hline 
  $P_0+P_2$  & 0.5 & $10^{-3}$ & 4 & $4.6\%~~(0.310)$ & $5.2\%$ (0.253)  \\ \hline 
  $P_0+P_2 + P_4$  & 0.5 & $10^{-3}$ & 4 & $3.4\%~~(0.253)$  & $5.1\%~~(0.253)$  \\ \hline 
  $P_0+P_2$  & 1 & $10^{-3}$ & 15 & $2.8\%~~(0.310)$ & $4.7\%~~(0.152)$   \\ \hline 
  $P_0+P_2$  & 1 & $10^{-4}$ & 4 & $4.8\%~~(0.310)$ & $10.1\%~~(0.124)$  \\ \hline 
\end{tabular}
\label{summarytable}
\end{table}


\section*{Acknowledgments}
The authors are exceedingly grateful to Dida Markovic and Alkistis Pourtsidou for useful discussions. We would also like to thank the anonymous referee for their suggestions and critiques. BB acknowledges support from the Swiss National Science Foundation (SNSF) Professorship grant No.170547.m 
KK and HAW is supported by the European Research Council through 646702 (CosTesGrav). KK is also supported by the STFC grant ST/N000668/1 and ST/S000550/1. 
\noindent 

\appendix


\section{Comparison of {\it N}-body and COLA} \label{app:colanbodycomp}
\begin{figure}[H]
\centering
  \includegraphics[width=12cm,height=8cm]{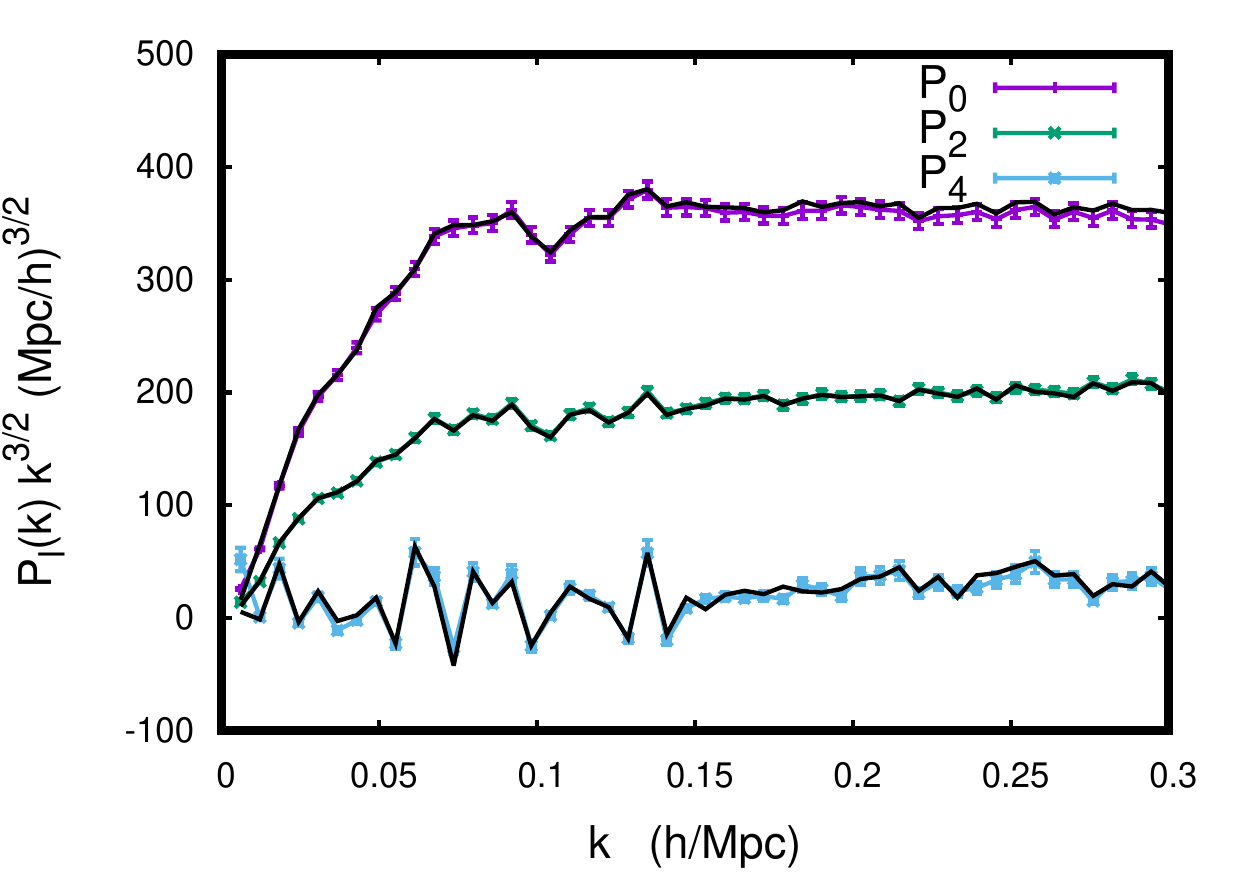}  
  \caption[]{Comparison of $P_\ell(k)$ in COLA and full {\it N}-body for one of the realisations used in this analysis. We have used exactly the same initial conditions in both simulations. The solid black lines shows the {\it N}-body result and the error bars displayed on the COLA results represent $2\%,2\%$ and $20\%$ for $P_0$,$P_2$ and $P_4$ respectively. The results shown are for $z=1$.}
\label{colacomparison}
\end{figure}

In this appendix we present a comparison of real and redshift space matter power spectrum of the halo distribution from full {\it N}-body to those obtained using approximate COLA simulations. COLA simulations are computationally much cheaper than doing full {\it N}-body simulations, but they are not exact and it's therefore important to ensure that the results are in agreement. This is  especially important when it comes to halos. Having a too low force resolution (low $N_{\rm grid}$ compared to $N$ particles in the simulation) or using too few time-steps in a COLA simulation can easily bias the halo population both in terms of abundance and halo properties. For a study on this see \cite{2016MNRAS.459.2327I}, but note that the results depends sensitively on simulations parameters like the boxsize, the number density and also on the halo finder used. We found that using a simple FOF halo finder gave the best agreement (apposed to using for example \texttt{Rockstar} which also takes into account velocity information to locate halos).
\newline
\newline
For one of the COLA realisations used in this paper we output the initial conditions and ran a full {\it N}-body simulation using \texttt{RAMSES} \cite{2002A&A...385..337T}. We computed FOF halo catalogs and estimated $P_0,P_2$ and $P_4$ for a subsample of the halos with number density $n = 10^{-3}(h/{\rm Mpc})^3$.  The comparison at $z=1$ can be seen in Fig.~\ref{colacomparison}. $P_0$ is shown to agree to $\sim 1-2\%$ down to $k=0.300\,h/{\rm Mpc}$, $P_2$ agrees to $\sim 1\%$ and for $P_4$ the scatter is quite big, but the overall agreement is typically $\sim 5-10\%$.

\section{Testing the Bias Model} \label{app:bias}
\begin{figure}[H]
\centering
  \includegraphics[width=12cm,height=8cm]{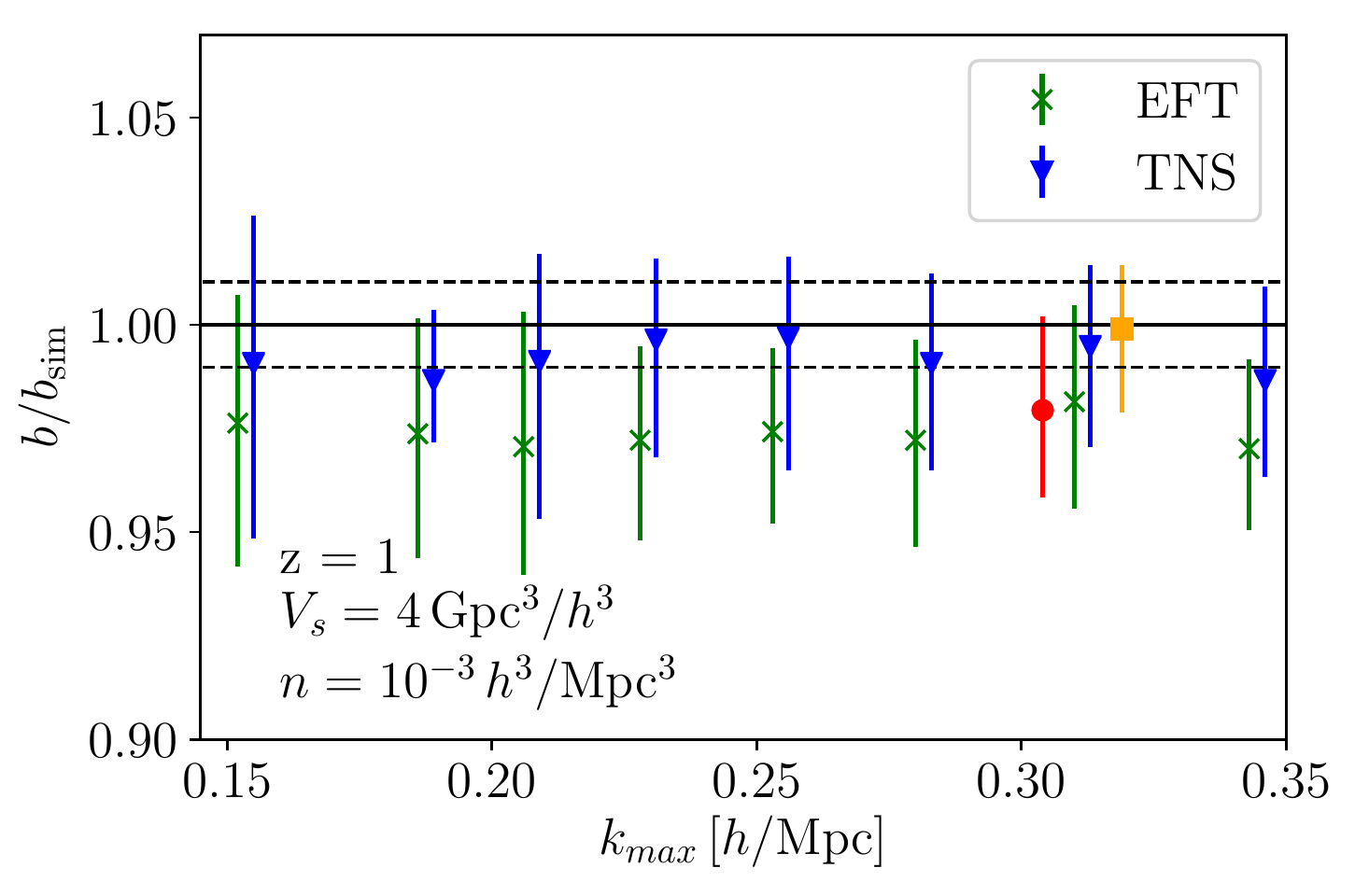}  
  \caption[CONVERGENCE]{ Redshift space halo results at $z=1$ with $V = 4 {\rm Gpc}^3/h^3$ taken as the redshift bin volume and using a halo number density of $n=10^{-3} h^3/{\rm Mpc}^3$ both in selecting the halo catalog and in the analytic covariance matrix used in the analyses. We show the mean value of $b/b_{\rm sim}$ as a function of $k_{\rm max}$ using the TNS (blue triangles) and EFTofLSS (green crosses) models with the marginalised $2\sigma$ error bars. The dashed lines indicate the $2\sigma$ errors on the measurement from simulations. The green and blue points come from analyses where only $P_0$ and $P_2$ were used while the orange square and red circle indicate the analysis including $P_4$ at $k_{\rm max,4} = 0.186h/{\rm Mpc}$ and $k_{\rm max,4}=0.310h/{
  \rm Mpc}$ for the TNS and EFTofLSS model respectively. Note the EFTofLSS and TNS models give biased (at the $2\sigma$ level) estimates of $f$ above $k_{\rm max} =0.310h/{\rm Mpc}$.}
\label{bias1}
\end{figure}

\begin{figure}[H]
\centering
  \includegraphics[width=12cm,height=8cm]{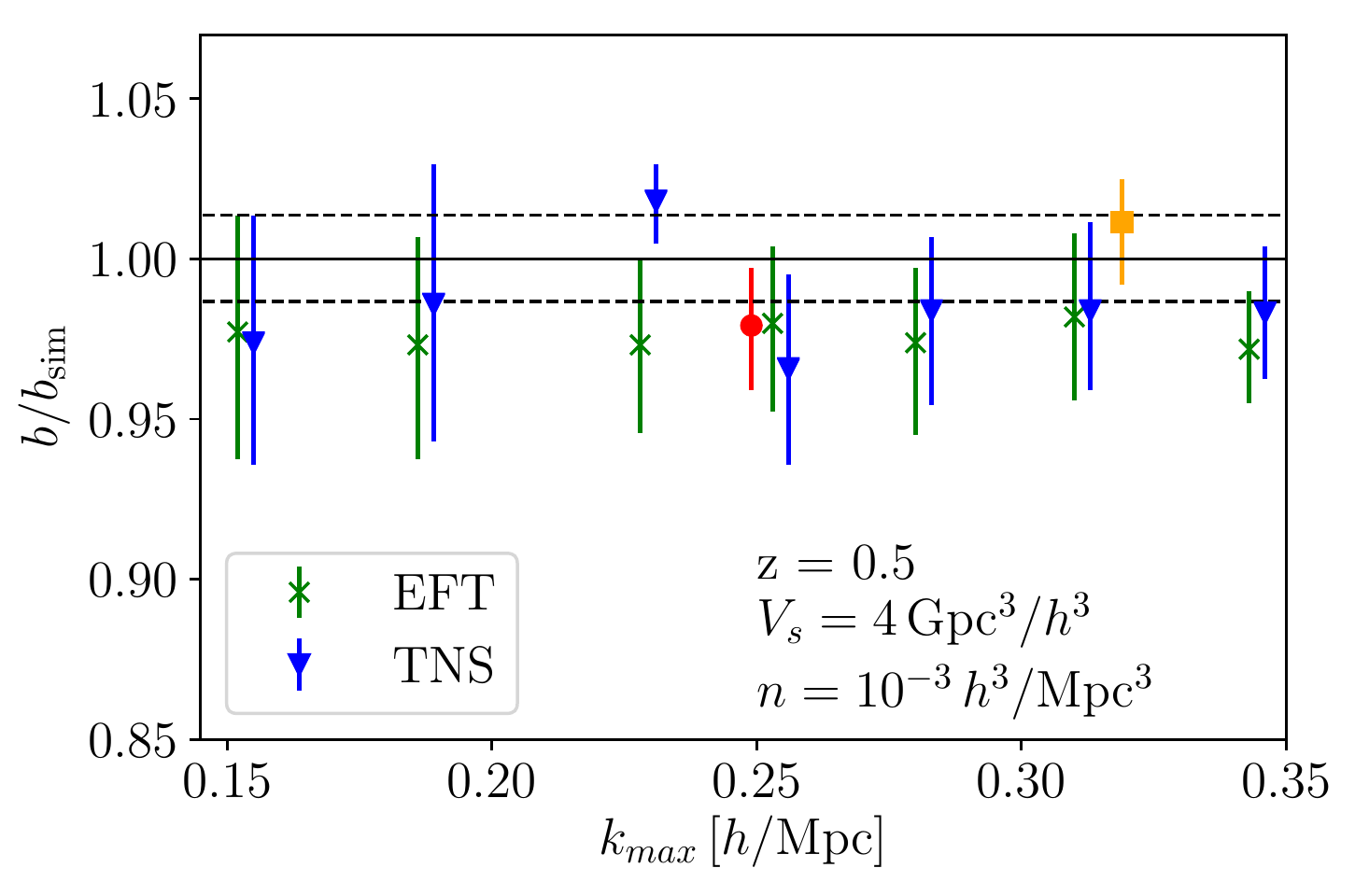}  
  \caption[CONVERGENCE]{Same as Fig.~\ref{bias1} but at $z=0.5$. Again the orange square and red circle  indicate the analysis with $P_4$ included at $k_{\rm max,4} = 0.253h/{\rm Mpc}$ for both models. Note the EFTofLSS model gives biased (at the $2\sigma$ level) estimates of $f$ above $k_{\rm max} = 0.253h/{\rm Mpc}$ while the TNS above $k_{\rm max} =0.310h/{\rm Mpc}$. }
\label{bias2}
\end{figure}

\begin{figure}[H]
\centering
  \includegraphics[width=12cm,height=8cm]{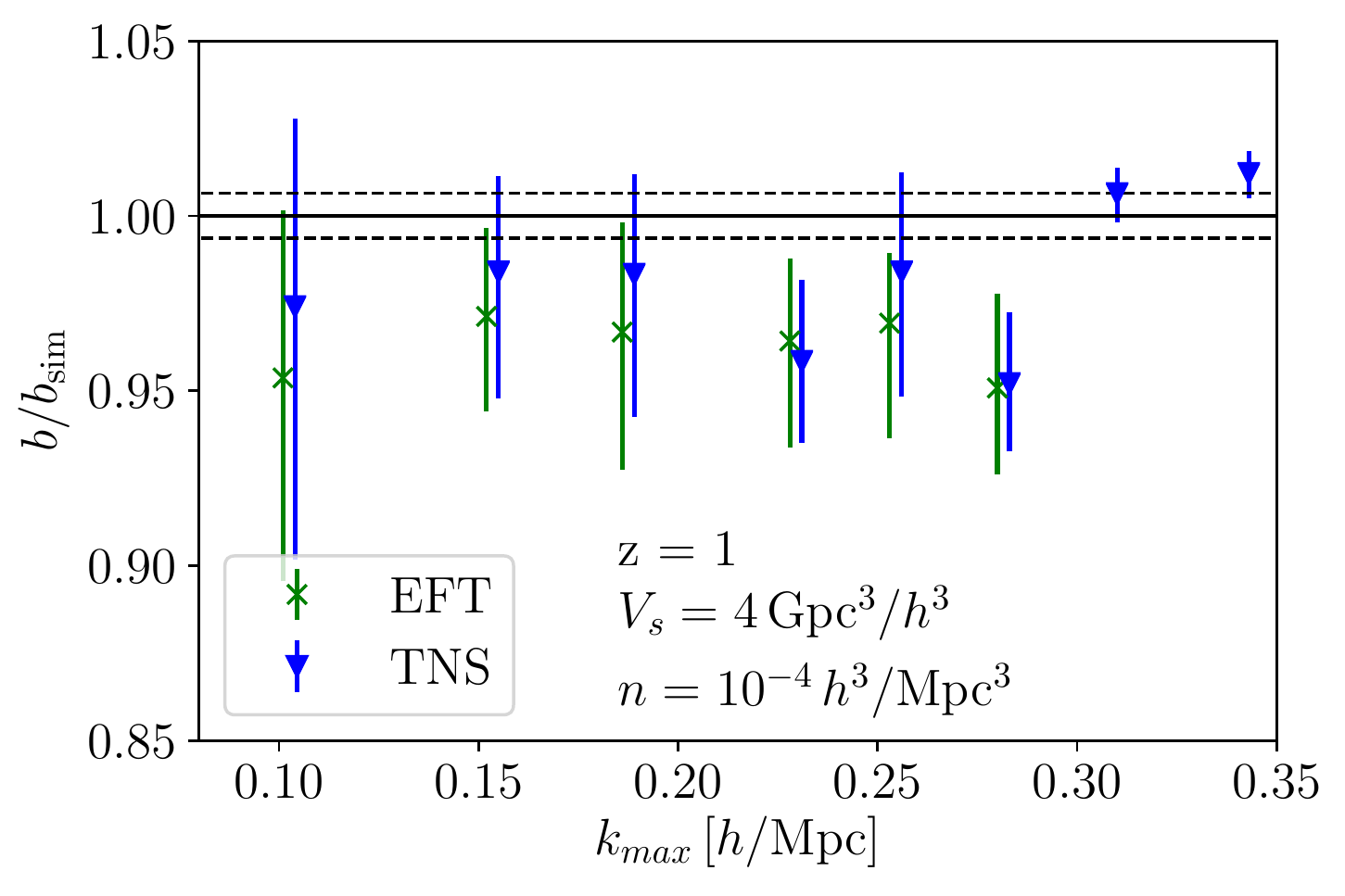}  
  \caption[CONVERGENCE]{Same as Fig.~\ref{bias1} but with $n=10^{-4}h^3/{\rm Mpc}^3$. Note the EFTofLSS model gives biased (at the $2\sigma$ level) estimates of $f$ above $k_{\rm max} = 0.124h/{\rm Mpc}$ while the TNS above $k_{\rm max} =0.310h/{\rm Mpc}$.}
\label{bias3}
\end{figure}

In this appendix we provide an additional test for the models, specifically we check at which scales the TNS and EFTofLSS give biased estimates of the linear bias $b_1$. This value can be measured from the halo and matter simulation spectra in the large scale limit. We would expect a full model of tracer bias to maintain this value for $b_1$ even when fitting to the small scales since additional bias degrees of freedom should capture non-linear bias effects independent of $b_1$. In this way, if the models predict a value for $b_1$ that does not match the 'fiducial' $b_1$  it is an indication that the bias model is failing and additional modelling is required. One may also expect such a failing to be strongly correlated with resulting biased estimates of cosmological parameters. 
\newline
\newline
In Fig.~\ref{bias1} we show the mean value of $b_1$ from the MCMC analyses at $z=1$ with $V = 4 {\rm Gpc}^3/h^3$ and  $n=10^{-3} h^3/{\rm Gpc}^3$ at varying $k_{\rm max}$. We also show the $2\sigma$ error bars from the analyses as well as dashed lines representing the $2\sigma$ errors from the simulation measurements.  We find that both the TNS and EFTofLSS model do not produce biased values of $b_1$ for $k\leq 0.343h/{\rm Mpc}$. The inclusion of the hexadecapole also does not bias the recovered value of $b_1$, shown as orange and red dots on the plot. Similarly, Fig.~\ref{bias2} shows the $z=0.5$ case. Again, both models do not show any biasing of the value of $b_1$. 
\newline
\newline
Finally, when we consider more highly biased tracers (more massive halos) the recovered value of $b_1$ does indeed move away from its measured value. These results are shown in Fig.~\ref{bias3}. We find that both of the models prefer lower values of $b_1$ that are more than $2\sigma$ away from the measured value when we consider scales $k>0.186h/{\rm Mpc}$. This is reflected in Fig.~\ref{halo13} where we see an early biasing of the recovered value of $f$ by the EFTofLSS model. 

\bibliography{mybib}{}

\end{document}